%% file: cola2022.tex
\documentclass[preprint,12pt]{elsarticle}

\usepackage{amssymb}

\usepackage{comment}
\usepackage{xcolor}
\usepackage{url}
\usepackage{multirow}
\usepackage{longtable}
\usepackage{pdflscape}

\journal{Computer Languages}

\begin{document}

\begin{frontmatter}



\title{Requirements Development for IoT Systems\\ with UCM4IoT}


\author[ru]{Paul Boutot}
\ead{pboutot@ryerson.ca}

\author[ru]{Mirza Rehenuma Tabassum}
\ead{mirza.tabassum@ryerson.ca}

\author[ru]{Abdul Abedin}
\ead{abdul.abedin@ryerson.ca}

\author[ru]{Sadaf Mustafiz\corref{cor1} }
\ead{sadaf.mustafiz@ryerson.ca}

\affiliation[ru]{organization={Department of Computer Science, Toronto Metropolitan University},
            addressline={245 Church Street}, 
            city={Toronto},
            postcode={M5B 1Z4}, 
            state={ON},
            country={Canada}}
            
\cortext[cor1]{Corresponding author}

\begin{abstract}
The engineering of IoT (Internet of Things) systems brings about various challenges due to the inherent complexities associated with such adaptive systems. Addressing the adaptive nature of IoT systems in the early stages of the development life cycle is essential for developing a complete and precise system specification. In this paper, we propose a use case-based modelling language, UCM4IoT, to support requirements elicitation and specification of IoT systems. UCM4IoT takes into account the heterogeneity of IoT systems and provides domain-specific language constructs to model the different facets of IoT systems. The language also incorporates the notion of exceptional situations and adaptive system behaviour. Our language is supported with a textual modelling environment to assist modellers in writing use cases. The environment supports syntax-directed editing, validation of use case models, and requirements analysis. The proposed language and tool is demonstrated and evaluated with two case studies: smart store system and smart fire alarm system. 

\end{abstract}




\end{frontmatter}



\input{introduction}

\input{background}

\input{usecaseml}

\input{process}

\input{toolsupport-v2}

\input{casestudy_ss}

\input{casestudy_sfa}

\input{comparison}
\input{discussion}

\input{relatedwork}

\input{conclusion}



 \bibliographystyle{elsarticle-num} 
 \bibliography{cola2022}






\end{document}

%% file: introduction.tex
\section{Introduction}

The Internet of Things (IoT) is bringing about a rapid evolution in the engineering of software systems. IoT is a heterogeneous set of interconnected things (machines, objects, people, animals), middleware, and software (smart) systems. IoT is revolutionizing application domains from consumer and commercial to industrial and infrastructure~\cite{atzori2010, reggio2020-survey}.

The \emph{things} constituting an IoT system have common goals and need to interact and cooperate to fulfill the functionalities of a smart system. The need to communicate with and control physical devices assigns new characteristics to such systems. The many facets of IoT - software, hardware, network, and environment - have to be taken into consideration when specifying the system interactions. Moreover, the adaptive nature of IoT systems introduces complexities and challenges that need to be addressed at the early stages of development. Discovering and documenting the complex interactions between the different participants or entities in such a system is critical for successfully developing the system. Potential exceptional situations need to be identified and possible adaptive behaviour also need to be explored. Deviations from the expected behaviour that are not identified during requirements elicitation might eventually lead to an incomplete or ambiguous system specification during analysis, and ultimately to an implementation that lacks certain functionality, or even behaves in an unreliable way. While the complexity of the requirements is recognized, there is still no established means of capturing the various elements and describing the complex interactions between the different dimensions of such systems.

In requirements engineering, use cases are an established means of discovering and detailing the system requirements. However as reported in \cite{parachuri2014}, ``structural defects can occur when use cases are written without following guidelines". A maximally constrained language can help designers define precise and complete requirements.

Previously, we have proposed a domain-specific requirements elicitation environment for IoT systems. We have taken inspiration from the exceptional use cases approach presented in \cite{smkd2005} and adapted it for IoT. Our language, UCM4IoT (\textbf{U}se \textbf{C}ase \textbf{M}odelling for \textbf{I}nternet \textbf{o}f \textbf{T}hings) provides a template as well as explicit guidelines for developing textual use case models. It facilitates discovery and specification of exceptional scenarios and adaptive mechanisms. It extends traditional use case models with IoT-specific elements. The language is supported by a modelling environment that enables modellers to write and validate their UCM4IoT models. The use of an extended use case diagram is proposed for summarizing the discovered requirements. 

This paper is an extension of ~\cite{boutot2021_sam}.
It introduces new features in UCM4IoT along with an additional IoT case study, smart fire alarm system. We present an evaluation of our work using two case studies. We have extended the language and approach with the following concepts and constructs: 1)services and modes of operation, 2) multiplicities of actors, 3) global exceptions, 4) specification of internal system processing steps and further categorization of the internal steps, and 5) explicit definition of exceptions in order to differentiate between the definition and occurrence of an exception. The modelling environment has been extended with new syntax and validation checks to support all new language extensions. We have also added tool support for generation of mode summary tables, handler summary tables as well as a new view of the exception summary table (use case view, in addition to the previous global view). Moreover, we provide more details on the process, approach, and tooling, and have also expanded the related work section with a detailed comparison table.

This paper is structured as follows: Section~\ref{sec:background} provides essential background. Section~\ref{sec:language} presents our modelling language and Section~\ref{sec:process} describes a process to support use of the language. Section~\ref{sec:tooling} discusses the modelling environment. Section~\ref{sec:casestudy-ss} and Section~\ref{sec:casestudy-sfa} demonstrates the application of our language on two IoT case studies. Section~\ref{sec:evaluation} evaluates our approach using the case studies. Section~\ref{sec:discussion} discusses potential extensions of the current work. Section~\ref{sec:relatedwork} includes related work and Section~\ref{sec:conclusion} concludes the paper.

%% file: background.tex
\section{Background}
\label{sec:background}


\subsection{Use Cases}

Use case modelling is an established means for elicitation and specification of system requirements~\cite{jacobson1992, cockburn2001, nasr2002}. Use cases 
are written in plain text, hence making it well-suited for communication with stakeholders.  Use case modelling helps in the discovery process by bringing hidden requirements to the surface.
A use case represents a scenario or a description of system behaviour that is required to satisfy a user goal. A goal can be defined at different levels of abstraction: summary, user goal and sub-functional. Use case descriptions (often referred to as textual use cases) 
includes a \emph{main success scenario} for describing the system interactions leading to a successful outcome (of a goal) and an \emph{extensions block} for specifying alternate scenarios. Every goal is associated with a set of actors: primary (actor with goal on system), secondary (actors with which the system has a goal and creates value for other actors), or facilitator (actor used by a primary or secondary actor to communicate with the system). Each use case is identified with a name, scope (context), intention (of the primary actor), level, and multiplicity. Use cases are scalable, since they can be decomposed into other use cases. Textual use cases are not part of the Unified Modelling Language (UML) standard. 
UML use case diagrams give a summarized view of the use cases (scenarios and system goals), relationships among the goals, and associations with external actors (systems, components, or human agents interacting with the system). 
These diagrams are usually complemented with textual use cases that contain essential information on the interactions. Such use case descriptions are much more appropriate for requirements elicitation over more formal diagrams such as UML activity or state diagrams. In our work, we use the well-defined textual use case template proposed by Fondue~\cite{fondue2000}, a UML-based software development method for reactive systems. A use case model in Fondue comprises of a set of textual use cases and a use case diagram. In comparison to Cockburn's loosely-defined use cases, the strict guidelines and template offered by Fondue make it easier and less confusing for novice users to write complete and precise use cases. 



While failure scenarios have been informally discussed in Cockburn's use cases, the notion of exception handling in use cases was introduced in the work on exceptional use cases~\cite{smkd2005}. Exceptions in use cases refer to exceptional situations that interrupt the normal flow of interaction and may require special handling to be carried out. It should be noted that requirements-level exceptions are not the same as design-level exceptions~\cite{wirfs-brock2002}. Exceptions internal to the system would prevent the system from fulfilling the user goal. Exceptions may also occur in the environment and change the context of the interaction. In both cases, the exceptions need to be detected and addressed by defining handling mechanisms. A special type of use case, referred to as a handler use case, is used to specify the exceptional interactions required to handle an exception. A UML profile for use case diagrams is provided to support creation of diagrams with exceptions and handlers. 

The use case model is a representation of the problem space, not the solution space. UML sequence diagrams are also used for interaction modelling, however they are more appropriate for requirements analysis and specification, but not for elicitation. 
It should be noted that use cases are not the same as user stories used in agile methods.
It is difficult to assess if the stories reflect the true reality and whether all essential requirements have been identified. For complex systems (such as, IoT systems), use cases, which give a detailed, clear and unambiguous description of the system behaviour are more appropriate for requirements development. 



\subsection{IoT Architectural Reference Model}

The IoT Architectural Reference Model (ARM)~\cite{iotarm2016} was established within the European research project, IoT-A, as a step towards standardizing a reference model for the IoT domain. In addition to introducing several views (context, functional, information, etc.), a domain model is provided which encompasses the common concepts of IoT systems along with the associations among the entities. 
The metamodel includes two kinds of users, \emph{human users} and \emph{digital artefacts} (software applications, agents, or services). A user invokes \emph{services} and interacts with \emph{physical entities}. A physical entity may be a type of \emph{device} (sensor, actuator, or tag) along with associated software (network or on-device resources). Notions of virtual entity and augmented entity (composition of a virtual and physical entity) are also part of the domain model.

%% file: usecaseml.tex
\section{Use Case Modelling Language}
\label{sec:language}
 There are many facets of IoT systems: software, hardware, network, and the environment in context. \emph{Software} is the core entity behind smart systems. The software controls the system and facilitates the interactions between the system and the environment, hardware and network. \emph{Hardware} includes physical entities or devices that are part of an IoT system and have to be detailed out at the requirements stage to ensure that all interactions with the hardware are considered. The \emph{environment} brings in a lot of complexity as well as uncertainty. An IoT system has to be designed to be aware of the uncertain nature of the environment in which the system is operating. The environment is made up of human users, physical entities, as well as all smart systems/objects/devices that may interact with the system via the Internet. The \emph{network} is the glue that brings together the environment and the system as well as facilitates communication between the various \emph{things} interacting with the system under development.

Unlike traditional software applications, IoT systems should be designed and developed by taking the different facets of IoT into consideration from the early stages of development. Eliciting IoT requirements is a challenge since typically requirements development techniques do not take into account these aspects. In this section, we propose a domain-specific requirements elicitation language for IoT systems, UCM4IoT. UCM4IoT aims to facilitate the elicitation process by bringing hidden requirements to the surface with the use of IoT-specific constructs. UCM4IoT allows specification of the requirements with a set of textual use cases (a UCM4IoT model). A use case diagram language with IoT-specific extensions is also proposed for summarizing the requirements. 

As part of the requirements development phase, a domain model (a structural view of the system) should be developed along with the use case model (outside the scope of this paper).


\subsection{Actors in UCM4IoT}

Use cases elaborate on the interactions of the system with the participating primary and secondary actors. The possible types of actors that are part of an IoT system are discussed here. The definitions of the types are according to the IoT Architectural Reference Model~\cite{iotarm2016}.
    
    \noindent\textbf{Human User:}  
    The user of the system is typically a human user. This is a type of primary actor that initiates or drives system interactions. Human users may also be secondary actors, i.e. actors participating in fulfilling the goal of a primary actor.
    
    \noindent\textbf{Physical entity:} These entities are part of the physical environment and play a role in satisfying a user's goal. A physical entity can be any object in the environment, for instance, living entities (humans, animals), moving objects (vehicles), immobile objects (buildings, stores, factories), electronics (appliances, computer, mobile devices), equipment, personal items (clothes, shoes), edible items (food). A physical entity has one or more associated devices.

    \noindent\textbf{Leaf Device:} This is a special kind of IoT element associated with a physical object. The devices form a core part of a smart system and can be of three basic types: sensors, actuators, and tags. We refer to a leaf device as a \emph{device} in this paper.
    \emph{Sensors} are used to monitor physical entities and provide information or data on the entity. Special types of sensors, known as \emph{readers}, can also be used to identify an entity.
    \emph{Actuators} are used to change the physical state of an entity (e.g. switching on a robotic vacuum or moving the direction of the vacuum). 
    \emph{Tags} are attached to entities to enable identification of the object (e.g., a barcode label or a RFID tag). The identification is carried out with the use of \emph{readers}.
    A device may be embedded within a physical entity (e.g., an actuator in a robotic vacuum, a sensor within a human body), attached to an object (e.g., a barcode tag on a store item), placed close to an object (e.g., a temperature sensor placed near a plant), or placed in the operating environment (e.g., a sensor placed in a room to detect humidity levels, a face-recognition enabled camera). 
    
    \noindent\textbf{Software:} Typically, communication with IoT devices is carried out with the use of a software application. Such an application is usually mediated by a human user. However, it might be running automatically to monitor a smart environment by gathering and analyzing data (e.g., monitoring temperature in a smart farm) or to control an environment (e.g., to automatically start a sprinkler system at a given time).

Physical entities in the IoT world typically need a virtual entity. The physical along with the virtual entity represents the ``thing" in the Internet of Things. At the use case level, the goal is to elicit requirements based on the problem space or the application domain. Concerns related to the solution space is not taken into account at this stage. Hence, the virtual aspect is not considered in our use cases. A physical entity identified during requirements elicitation will need to be associated with a corresponding virtual entity in the design phase.

Having specific types of actors constrains the modeller to IoT-specific concepts when developing the use cases. The system can interact with actors of specific types and the interactions in the use cases can be validated to ensure that only permitted types of actors are part of the scenarios. 

\subsection{Exceptions and Handlers in UCM4IoT}

Abnormal or exceptional situations are identified with defined \emph{exceptions} in the use case model. Such exceptions identify  
situations that occur unexpectedly, interrupt the normal flow of interaction, and require the system to tolerate and possibly adapt to the new circumstances. Unless these exceptions are discovered and documented, no special handling will be incorporated in the system, leading the system to have missing functionality and to behave in unexpected ways in such situations.

In this work, we classify exceptions that can occur in an IoT system to be of the following types: hardware, software, network, and environment. A \emph{hardware exception} denotes abnormal behaviour originating in a physical entity or device in the system (e.g., sensor failure, actuator failure). A \emph{software exception} represents unexpected behaviour of a software system or subsystem participating in fulfilling a user's goal (e.g., credit card system down). This does not refer to exceptions in the software system under development, which come up later in the design and implementation phase. At the requirements development stage, the focus is on ``what" the system should be doing, not on ``how" it should be done.
A \emph{network exception} occurs when a system or \emph{thing} is unreachable, potentially due to network issues (e.g., fire station alarm unreachable, robotic vacuum unreachable).
An \emph{environment exception} represents an unexpected situation occurring in the external environment in which the system is operating. Such an exception would typically interrupt and suspend or terminate a user's goal (e.g., a fire hazard in a smart store). 

An exception can occur in more than one use case. Each occurrence of an exception is regarded as a different instance of the same exception - the origin or source of the exceptional situation is not the same. However, some exceptions (for instance, a fire hazard) can occur anytime and impact many use cases. In UCM4IoT, these exceptions are classified as global exceptions. While an environment exception typically would be identified as a global exception, other exception types can also be identified as global. As an example, a network exception (no internet connectivity) in the smart fire alarm system, would impact several user goals at the same time.

The environment is not under our control. Unaddressed uncertainties arising in the environment can lead to incomplete requirements and eventually missing functionalities in the target system. In the case of hardware, all hardware are prone to failure. Depending on the quality of service requirements of a system, fault tolerance mechanisms are incorporated at the design and implementation levels to deal with hardware failures.

In an adaptive system, an exceptional situation identified during the execution of a user's goal would require the system to handle and attempt to recover from the situation in the most feasible way. The adaptive behaviour of the system needs to be detailed by specifying the system interactions in such situations. In our language, we use a special type of use case, a handler use case, to specify the steps to be completed to handle a certain exception. When an exceptional situation occurs, the base behaviour is put on hold and the interactions in the handler are initiated. A handler can temporarily take over the system interaction, perform some compensation tasks, and then return to the normal interaction scenario. In cases when the handler cannot help recover from the exception, the exception would cause the goal that was interrupted to fail. At times, a new actor, referred to as an \emph{exceptional actor}, may participate in the handling actions. These actors are introduced in handler use cases.

\subsection{Services and Modes} 
As defined by The Open Group, a service is a logical representation of a set of activities that has specified outcomes, is self-contained, may be composed of other services, and is a ``black-box" to consumers of the service~\cite{og-soa-2016}. 
Services are reusable components of a system that are independent and loosely coupled ~\cite{sommerville2015software}.

While developing a system, designers often try to reuse existing services that can be used to fulfill some goals of the system. After the requirements elicitation phase, designers identify existing services from service repositories to meet business requirements, and additionally develop required services for the system.
However, often services are reused that do not meet the business requirements fully. Reusing a service that partially matches the requirements results in an inaccurate system \cite{bano2013makes}. Services must be aligned with the requirements, not vice versa.

Use cases describe actors and system’s interaction without explaining the internal technical details of that system. As use cases lay out goals of the system, and services are developed to meet user goals, the use cases are used as a tool for describing operational activities in service oriented architecture \cite{sasikalastudy}. Goals are specified from the perspective of the users of the system, while services are specified from the system's perspective. 

A system can operate in different modes in different circumstances. A mode is defined by the set of services that is exposed by the system while operating in a certain mode\cite{mustafiz2009drep}. On the event of an exceptional situation, a system may not be able to provide all services, or it may only be able to provide exceptional or emergency services. In such cases when future service provision may be impacted, the system would switch from a normal mode of operation to a degraded, restricted, or emergency mode of operation~\cite{mustafiz_serene2008}.

UCM4IoT accommodates different modes of operation in the use cases. The categorization of modes in UCM4IoT is based on the definition of modes proposed in \cite{mustafiz_serene2008}. A system operates its day to day business in the \texttt{normal} mode. In case of an emergency, the system may need suspend normal operation and offer some special services to address and handle the scenario by switching to an \texttt{emergency} mode. There may be special circumstances when the system can only offer limited services (a subset of the normal services) as well as required emergency services, putting the system in a \texttt{restricted} mode. System can also operate in a \texttt{degraded} mode when certain services are unavailable or offered with reduced quality-of-service. 

There is an implicit connection between goals, services, and modes. Services are offered by the system to achieve the goals of actors, and different modes offer different sets of services. Hence, meeting goals of the system also depends on which mode the system is in at that time. In UCM4IoT, the default mode is the \texttt{normal} mode, hence all use cases, except handler use cases start in a \texttt{normal} mode. However, a mode switch can occur during the execution of a use case or handler use case. A step in the alternate or exceptional block may also trigger a mode switch. 
Multiple mode switches are also possible within a use case in UCM4IoT. Moreover, it may also be possible for a system to offer more than one normal mode of operation.

\subsection{Representing Exceptions, Handlers, and Modes in Use Cases}

Similar to a standard use case, a use cases in UCM4IoT includes a header with details on the name, scope, level (summary/user-goal/sub-function), intention, and multiplicity. Each use case has a set of associated actors categorized as primary, secondary, or facilitator. The syntax for specifying actors is {Type::UniqueName} (e.g., \texttt{Human::Customer}, \texttt{Sensor::Weight}). The body of the use case is composed of the main scenario (outlining the normal system interactions leading to a success of the user goal) and an extensions block (outlining the alternative paths and/or exceptional cases). A step in an use case can be of the following types: interaction (send or receive requests between the system and actors), invocation (calls another use case), condition (an interaction occurring in the environment), internal (internal system processing on which the following steps are dependent or a timeout specification), or control flow (a step dictating the flow of interactions). The main success scenario as well as each extension block have an associated outcome: success (denoting a successful execution of a goal), failure (denoting a failed execution of a goal), degraded (denoting a partially successful execution of a goal), or abandoned (denoting voluntary termination of a goal by a user). 

Exceptions are defined in the extensions block with the type and name of exception (e.g., \texttt{EnvironmentException::FireHazard}). In case of global exceptions, \texttt{::global} is added to the exception name. The UCM4IoT use case template is shown with the use of an example in Fig.~\ref{uc:exit}. Exceptions can be associated with single steps or a block of steps. If a step (or a block) includes an invocation, the exception can also be raised during the execution of the invoked use case (see  Fig~\ref{uc:shopping}). Every handler is associated with the context (use case in which the exception appears) along with the exceptions,  which is explicitly stated in the header with the \texttt{contexts \& exceptions} clause (see example in Fig.~\ref{uc:fire}). 

Modes are specified in use cases with the \texttt{mode switch} keyword. If a use case in normal mode switches to restricted mode after an exception is raised, it can be specified using the syntax \texttt{mode switch: Restricted.} 
 
An exception can trigger a mode switch, then the handler for the exception operates in that switched mode.

\begin{figure}[tbh!]
    \centering
    \includegraphics[width=1.0\textwidth]{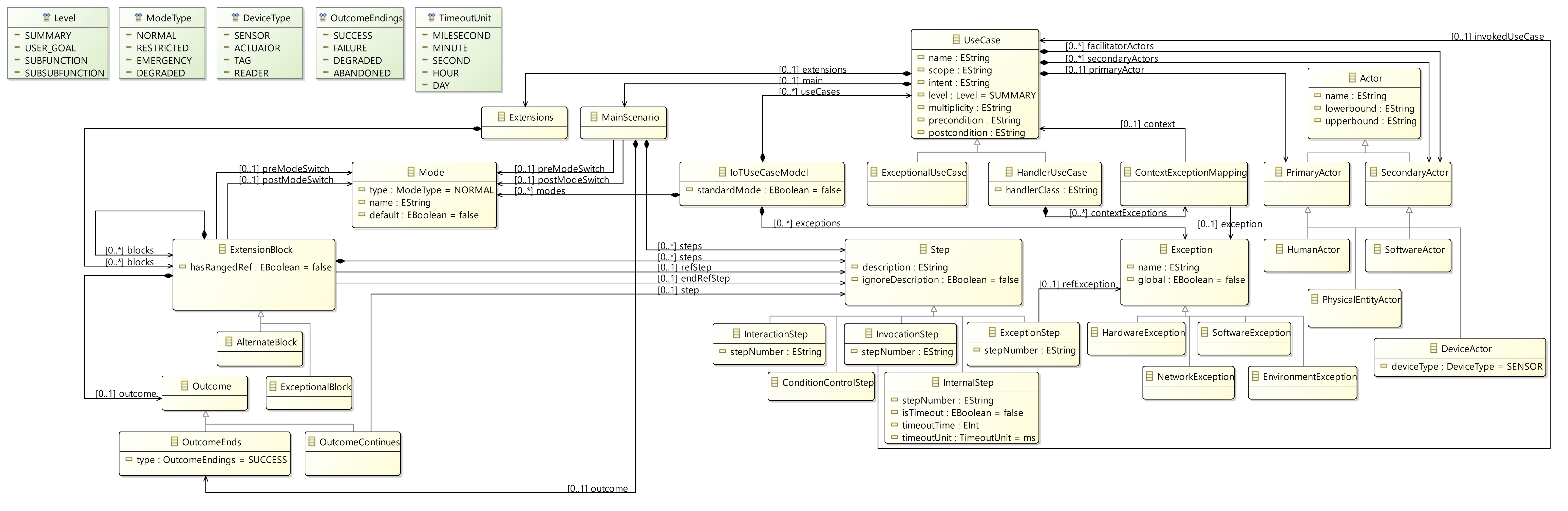}
    \caption{UCM4IoT metamodel.}
    \label{fig:metamodel}
\end{figure}

The relationship between handlers and use cases can be one of two types: 1) \emph{interrupt \& continue}: in this case, the user goal (including all sub-goals) is put on hold due to an exceptional situation - the use case execution resumes following the handling; or 2) \emph{interrupt \& fail}: when the user goal (including all sub-goals) fail due to an exceptional situation - the use case execution is terminated. 
The UCM4IoT metamodel is presented in Fig.~\ref{fig:metamodel}. 

For summarizing the outcome of the requirements elicitation process, we have adapted and used the exception-aware use case diagram proposed in \cite{smkd2005}. Standard UML use case diagrams have been extended with exceptions, handler use cases (a subclass of Use Case) and interrupt relationships (similar to Extend and Include relationships but specifically uses handler use cases as the source). The relationship between handlers and use cases are tagged with \texttt{<<interrupt \& continue>>} or \texttt{<<interrupt \& fail>>}. These dependencies are depicted in the extended use case diagram. We have used this UML profile with the IoT-specific elements and syntax introduced in UCM4IoT for specifying actors and exceptions (see example in Fig.~\ref{fig:ssucd}). The profile is not included here due to space constraints.

%% file: process.tex
\section{UCM4IoT Process}
\label{sec:process}

While the UCM4IoT language can be used for requirements elicitation according to the modeller's need, we provide here some recommended guidelines. 

\subsection{Inception} The requirements development process starts off with identifying business needs and understanding the problem and stakeholders. 
A problem statement followed by an initial set of hardware and software requirements (an informal natural language specification) is to be developed by gathering domain knowledge and analyzing the stakeholders and existing systems or documentation.

\subsection{Elicitation}   
The use case-based requirements elicitation process begins by identifying the major stakeholders of the system under development (referred to as \texttt{System} in the use cases) and by identifying their goals with the system. 
Based on the problem domain and communication with stakeholders, establish goals and services that the system needs to fulfill.
The primary actor is typically a human user, however in IoT systems, software and hardware systems may also initiate the system interactions. 
Based on the UCM4IoT actor types, the secondary and facilitator actors (humans, physical entities, software, devices) participating in a goal are identified. 
With the help of the UCM4IoT process guidelines, identify scenarios that lead to the success of the goals. 
The normal behaviour of the system in order to fulfill the goal is described in the main success scenario of the use case. Next establish all alternative behaviour that may arise in the system.

The language constructs should be used to identify different types of actors (human, physical entity, leaf device, and software). In this phase, exceptional behaviour also needs to be discovered based on the UCM4IoT language. Having the different types of actors make it easier to define granular interactions ensuring that every interaction between the system and the associated devices, physical entities, external software systems, or other human users are present in the use case. Relevant internal context steps or timeouts associated with each goal should also be specified. 

The system provides various services that fulfill one or more user goals. The services to be offered by the system need to be derived from the use cases established.

In addition to discovering actors and goals, the possible modes of operation supported by the system needs to be established. This includes normal modes as well as special modes (restricted, degraded, emergency) that the system may need to switch to in order to provide special or limited services. 

\subsection{Elaboration} This activity focuses on analyzing and elaborating the requirements. Once the normal behaviour is defined, each interaction in the main scenario needs to be revisited to explore possible exceptional situations that may arise. The following questions can be used to identify exceptional situations: \emph{Can this interaction step fail in any way? What might be possible sources of failure? Are there any exceptional situations that may arise due to problems with the hardware, software, or network? Can any external event originating in the operating environment change or fail the user’s goal?} Having explicitly defined types for the exceptions, makes modellers aware of the different aspects associated with a given interaction and assists them in identifying potential hardware, software, network, or environment exceptions that would otherwise not be discovered using standard use cases.

As the next step, handling mechanisms for each identified exception should be explored. The following probes can be used to elicit handlers: \emph{How should the system react and adapt in such an exceptional situation? What additional system functionality is required to enable the system to continue fulfilling the user’s goal in a such situation? Does the system need the help of an (exceptional) actor to handle the situation?} In the case of hardware exceptions, a possible handling mechanism would be to call a service person for repairing or replacing the device. Software exceptions are due to issues with an external system, and hence in such scenarios, there should be some compensation activity carried out by the system so that the system does not fail to fulfill the user’s goal. Network exceptions can be handled by periodically attempting to reconnect or by trying to communicate with the actor using some alternate means. To address environment exceptions, there may be a need to support such exceptional situations with added system functionality that was not part of the initial problem statement and desired system features.

The handlers should also be analyzed for exceptional behaviour. The handlers need to be documented. Designers should ensure that all outcomes are clearly defined in the model. UCM4IoT gives warnings and errors to assist in the completion of the use case model.

\subsection{Negotiation/Prioritization} Static analysis with UCM4IoT: The UCM4IoT tool should be used to generate the mode table, exception summary table, and handler summary table. Based on the use cases, UCM4IoT generates a summary of paths that result in an exception being thrown. The tool also generates summary information on handlers and their associated use cases. The criticality of use cases, exceptions and handlers can be determined based on the generated information and higher weight or priority can be attached to the corresponding functionality when design decisions are taken later on expected quality-of-service and fault tolerance measures to be used. 

\subsection{Specification} The UCM4IoT model (textual uses cases and the use case diagram), the accompanying summary tables and the IoT ARM domain model are all part of the requirements specification of the system. 

\subsection{Validation} 
Requirements validation involves examining the requirements model for inconsistency, omissions, and ambiguity. The use cases model needs to be validated to ensure that there are no inconsistencies in the use cases, for instance, missing or mismatched actor types, missing use cases or handlers, ambiguities in the definition or use of actors and exceptions.

The UCM4IoT modelling environment, described in Section~\ref{sec:tooling}, further assists developers in eliciting and defining use cases, exceptions, and handlers for IoT systems.

%% file: toolsupport-v2.tex
\section{Tool Support}
\label{sec:tooling}

The UCM4IoT environment assists modellers in developing requirements for IoT systems with textual use cases. The environment was built in Eclipse utilizing its IDE Plug-in support and  Xtext\footnote{\url{https://www.eclipse.org/Xtext/}} which supports the development of textual modelling languages. Xtend\footnote{\url{https://www.eclipse.org/Xtend/}} was used for building support for validation and summary information generation.

An UCM4IoT model consists of two types of elements: standard use cases and handler use cases. Both types require several clauses. Specifically, the environment enforces that the use cases include the scope, level, intention, multiplicity, primary actor, and main success scenario clauses. It also supports optional clauses that the two types of use cases can utilize. These include the extensions, secondary actor, facilitator actor, precondition, and post-condition clauses.
Additionally, the tool specifies that handler use cases require one other clause, namely, the \emph{context and exceptions} clause.
An UCM4IoT model also includes a header located at the top of the model. This header consists of a list of modes and exceptions used throughout the UCM4IoT model.

Our modelling environment provides support for syntax highlighting, refactoring, type-checking, cross-referencing, validation, and data generation. Specifically, the grammar specifies the keywords of the modelling language, the allowed types for actors and exceptions, and the linking of references to their respective declaration. 
Moreover, model validation and data generation are carried out in the backend. 
Figure~\ref{uc:shopping} shows an example of syntax highlighting and type-checking. Notably, UCM4IoT highlights all of the clauses and types. 

UCM4IoT limits all steps found in the main success scenario or the extensions to either an invocation, interaction, control-flow, condition, internal step, or exception. 
Extensions consist of extension blocks, which then contain steps or more extension blocks. Extension blocks represent alternative paths from a declared step in the use case. These blocks can be either exceptional or alternative blocks. The main difference is that exceptional blocks contain one step that throws an exception. The environment will throw an error if the modeller fails to conform to the provided grammar.

UCM4IoT also lets users define the different modes of the system along with the mode switches. The user lists the default mode and all other modes in the header of the UCM4IoT model. Afterwards, the user can reference these modes in the main success scenario or any extension block to specify when the system switches into these modes. The system can only switch into a mode at the beginning or end of the main scenario or an extension block.

The following outlines the features of our environment.

\begin{figure}[tbh!]
    \centering
    \includegraphics[width=0.75\textwidth]{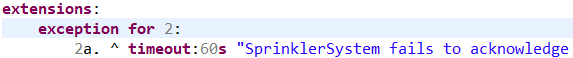}
    \caption{UCM4IoT Syntax: Internal step specifying a \emph{timeout}.}
    \label{fig:tool-timeout}
\end{figure}

\begin{figure}[tbh!]
    \centering
    \includegraphics[width=0.5\textwidth]{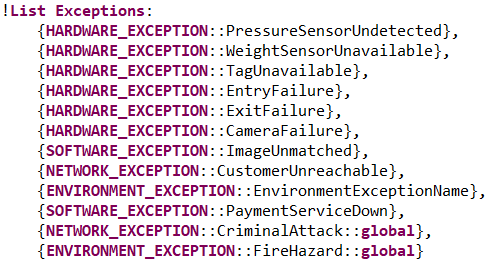}
    \caption{UCM4IoT Syntax: Exceptions defined in header.}
    \label{fig:tool-header-exceptions}
\end{figure}

\noindent\textbf{Step Ordering and Formatting:}
For the steps defined in the main success scenario and the extensions, UCM4IoT provides validation checks for their ordering and format. Any numbered step must follow the logical ordering of the previous step. Thus, this simplifies the creation of the main success scenario and the extensions as the environment notifies modellers when the sequence of steps is no longer correct. If a step does not follow the logical ordering of its previous step, UCM4IoT will throw an error message along with a list of possible step numbers that can occur after the last numbered step.
For invocation steps, our environment enforces that the modeller states the invoked use case of the step. Our environment cross-references the invoked use case, giving an error for use cases that are not defined. Likewise, exception steps require modellers to define the raised exception in the header (as seen in Fig.~\ref{fig:tool-header-exceptions}). Any exceptions that do not exist in the header result in the environment giving an error. Lastly, internal steps provide the modeller with the option to describe \emph{timeouts} in the system, denoted with a keyword and the amount of time 
(see Fig.~\ref{fig:tool-timeout}). The environment will give an error if the modeller incorrectly uses this feature.

UCM4IoT also enforces that all interaction steps found in summary-level and user goal-level use cases include the \emph{system} and an actor from the list of actors defined in the use case.
Lastly, UCM4IoT also features the ability to cross-reference steps. Any step mentioned in an outcome of a use case or the header of an extension block will be linked to an existing step found in the use case.

\noindent\textbf{Type Checking:}
 The tool checks to see if all actors in a use case match one of the possible valid types (human, software, device, or physical entity). Moreover, an actor can also be typed as a sensor, actuator, tag, or reader by the grammar; however, these types are considered as device actors by UCM4IoT. The environment provides an error for any actor or exception that forgets to include a type, for example \texttt{EntryGate} instead of \texttt{PhysicalEntity::EntryGate}. This clearly shows the type of the participating actor.
 UCM4IoT also enforces that all exceptions have their associated type defined.
 
\noindent\textbf{Multiplicity:}
UCM4IoT allows for modellers to specify the multiplicity of actors. The environment also checks the format of the multiplicity. Its lower bound must be lower than or equal to the upper bound. Moreover, the modeller may denote the range as "*" to indicate any amount. Syntax checks are carried out by the environment.

\begin{figure}[tbh!]
    \centering
    \includegraphics[width=1\columnwidth]{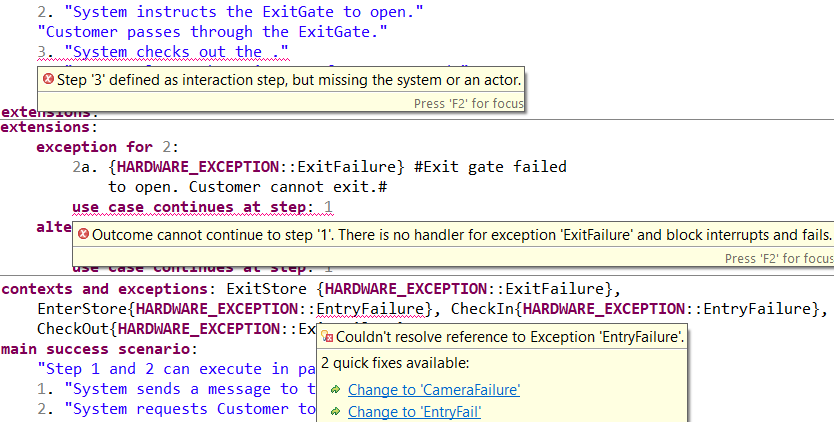}
    \caption{Sample validation checks.}
    \label{fig:tool-validation}
\end{figure}

\noindent\textbf{Exception Handling:}
For use case handlers, UCM4IoT checks if the defined contexts and exceptions are valid. If the exception does not appear in the use case context, the environment notifies the modeller about this issue (see Fig.~\ref{fig:tool-validation}: bottom).
Furthermore, UCM4IoT checks that exceptions that appear in a use case are actually handled in some use case handler. Otherwise, a warning is displayed (see Fig.~\ref{fig:tool-validation}: middle). Thus, the environment prevents the modeller from forgetting any unhandled exceptions.

UCM4IoT also requires modellers to define exceptions separately from their occurrence within use cases. An exception is defined in the header, and the modeller can choose to reference that exception within an exception step. It is necessary to distinguish between the definition and the occurrence of an exception so that multiple exception steps may refer to the same exception; otherwise, if two exception steps include the same exception, they would be referring to two unique definitions of exceptions that happen to have the same name. This difference is significant within the environment since the latter option prevents UCM4IoT from properly identifying handled exceptions. Since both exceptions are considered to be distinct, a handler use case attempting to refer to a raised exception would not be able to handle all occurrences of that exception. By requiring modellers to define exceptions in the header and raising them in use cases, UCM4IoT can link handled exceptions to their occurrence as all occurrences of an exception will be the same exception.

\noindent\textbf{Outcomes:}
UCM4IoT enforces that all main scenarios, exceptional blocks, and alternative blocks end in some outcome. Outcomes must either end in success, failure, degraded, abandoned, or continue with some other step. The main success scenario only has one outcome that always ends in success. In contrast, the extensions clauses can have many alternate steps, meaning that they can have many outcomes. This feature prevents the modeller from forgetting the outcome of any scenario. 
The tool also provides validation checks for the outcome of an exception block. Specifically, if there is no handler for a raised exception when an exception block continues to another step, the environment would display an error message stating that the use case cannot continue as the exception is never handled.

\noindent\textbf{Table Generation:}

\begin{figure}[tbh!]
    \centering
    \includegraphics[width=1.1\textwidth]{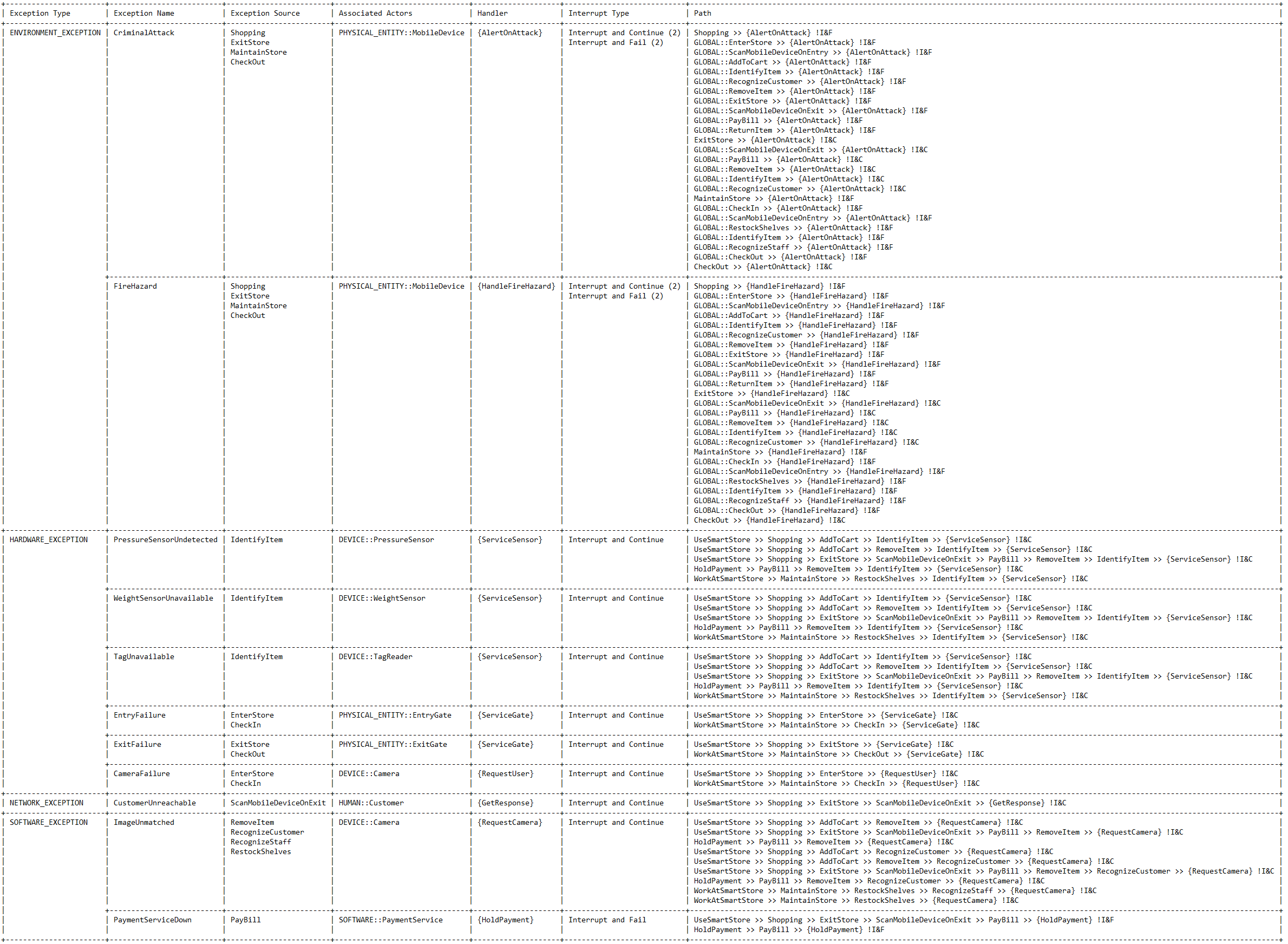}
    \caption{Sample exception summary table: global view.}
    \label{fig:tool-validation1}
\end{figure}

UCM4IoT provides the generation of exception summary tables for each use case within a UCM4IoT model. Any exception that could occur in a use case (including those found in any invoked use case) will be included along with the source use case of the exception, the handler that handles the exception, the list of possible situations that result in the exception occurring, and the actors that participated in the event that caused the exception to be thrown. A slice of an exception summary table is shown in Fig.~\ref{fig:tool-validation1}. It is possible to generate two different views: a global view and a use case view. A global view contains information about all of the exceptions within a UCM4IoT model. Comparatively, a relative view contains information about exceptions that can be raised when looking at a particular use case. This generated information can then be used for making inferences and carrying out application-specific analysis (discussed further in Sections~\ref{sec:casestudy-ss} and \ref{sec:casestudy-sfa}). Future iterations will allow users to import use cases from other UCM4IoT files, allowing support for larger-scale projects. 

\begin{figure}[tbh!]
    \centering
    \includegraphics[width=1\textwidth]{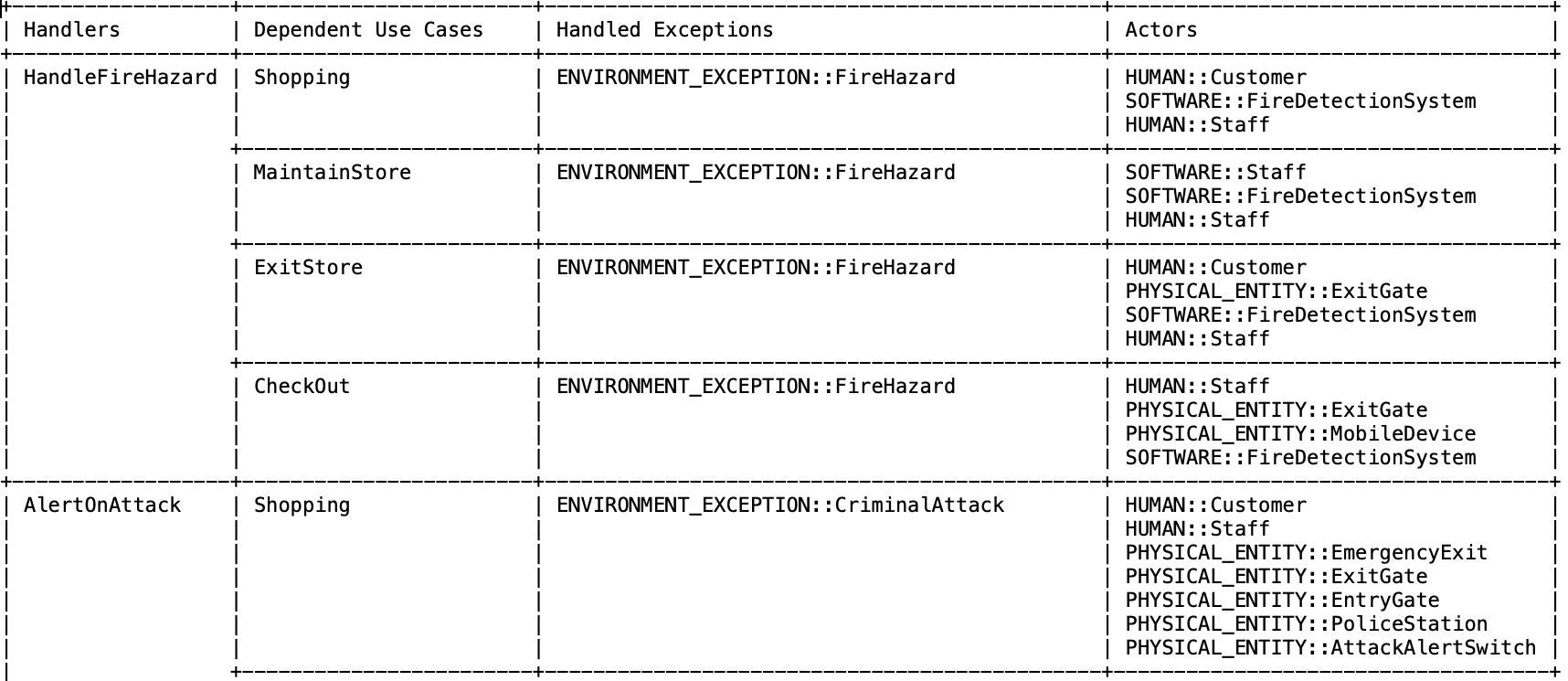}
    \caption{Sample handler summary table (slice).}
    \label{fig:tool-handlertable}
\end{figure}

UCM4IoT also supports the generation of handler tables for every handler within a UCM4IoT model. The table describes which use cases depend on the handler, the exceptions that it handles, and any exceptional actors that appear (denoted with a '*'). A snippet of a handler table is shown in Fig.~\ref{fig:tool-handlertable}. Moreover, the tool can generate a mode table listing names of all use cases along with mode switches. Fig.~\ref{fig:modeTable} shows the smart store system's mode switches. 

A demo video of our environment is available at\\ \url{https://www.cs.ryerson.ca/~pboutot/ucm4iot-new.html}.

\noindent\textbf{Interoperability:}

Moreover, the environment generates an exportable .xmi version of UCM4IoT 
to enable UCM4IoT models to be ported to other modelling tools. 

%% file: casestudy_ss.tex
\section{Case Study 1: Smart Store System}
\label{sec:casestudy-ss}

In this section, the use of the UCM4IoT language is demonstrated with a case study, namely the smart store system. 

\subsection{Smart Store System Overview}
The retail industry is being revolutionized in the IoT era with intelligent shopping, smart carts, and smart stores~\cite{li2017, karjol2018}. 

A smart store system is a brick and mortar, checkout-free, “walkout” store. 
\emph{Just walk out} technology (\url{https://justwalkout.com}), introduced by Amazon, allows customers to pick up items from the store and walk out without going through the usual checkout process. The system automatically processes payments by identifying items taken by the customers. 
Implementing this system requires automatic detection of various items in the store as well as recognizing customers purchasing the items. 

Users are required to complete the customer registration process and install the smart store mobile application on their mobile device prior to shopping at the smart store. Customers need to scan their mobile device (or alternatively, a pre-registered credit card) at the entry gate.
Cameras placed at the entry location take images of customers, and the system associates those images with respective customers. These images are used by the system later to recognize a customer inside the store. There are various sensors attached to the items' shelves to identify items. 
Cameras installed inside the store take images of customers while they are shopping and associate each customer with the items they took from the shelves. Customers scan their mobile devices or credit cards at the exit gate while leaving the store. Payment service deducts the bill, and then the system opens the exit gate. Customers can set up their preferred payment service beforehand and allow the system to process the payment automatically on exit. There are staff on premises for maintenance, assistance, and restocking purposes.

A set of informal hardware and software requirements (not included here due to space reasons) for the smart store system is initially derived based on the problem statement and domain knowledge. 
We then used the UCM4IoT approach to elicit and specify the system requirements. 

\subsection{Requirements Development with UCM4IoT}
The use case model for the smart store system has been developed using the guidelines elaborated in Section~\ref{sec:process}.

\subsubsection{Actors and Goals}
The smart store system has two primary actors, Customer and Staff, both of type Human User (\texttt{Human::Customer} and \texttt{Human::Staff}). The Customer's goal with the system is \texttt{Shopping} for items, and the Staff's is \texttt{Maintain Store}. 
 
The system also interacts with external software applications to help it achieve its goals, such as  \texttt{Software::PaymentService} for processing payments and \texttt{Software::FireDetectionSystem} for detecting fire hazards. 

For the customer's primary goal to have a successful outcome, a customer has to enter the store, purchase or return items, and then leave the store. Customers may be unable to shop in case of a fire or other hazards (e.g., criminal attack), leading to a failed outcome.
Participating actors have been categorized as per the UCM4IoT actor types. 
\begin{figure*}[tbh!]
\begin{minipage}{.5\textwidth}
    	\centering 	
		\includegraphics[width=1.07\textwidth]{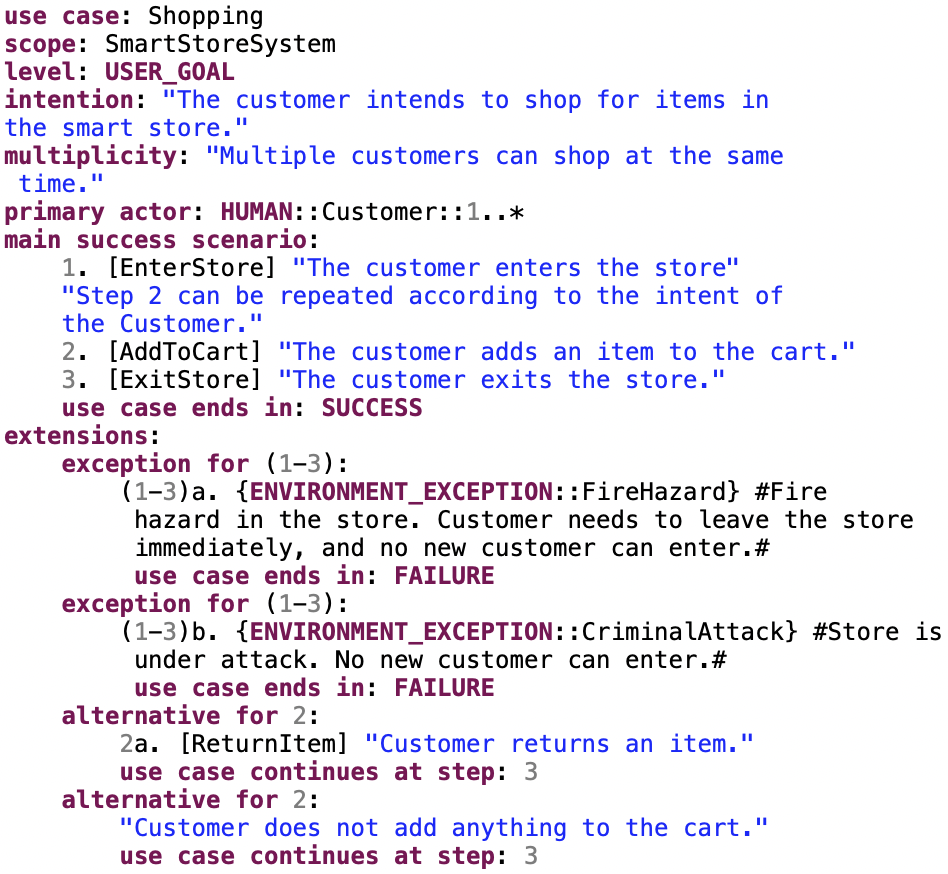}
		\caption{Smart store: Shopping use case.}
		\label{uc:shopping}    
\end{minipage}
\begin{minipage}{.5\textwidth}
	\includegraphics[width=1.1\textwidth]{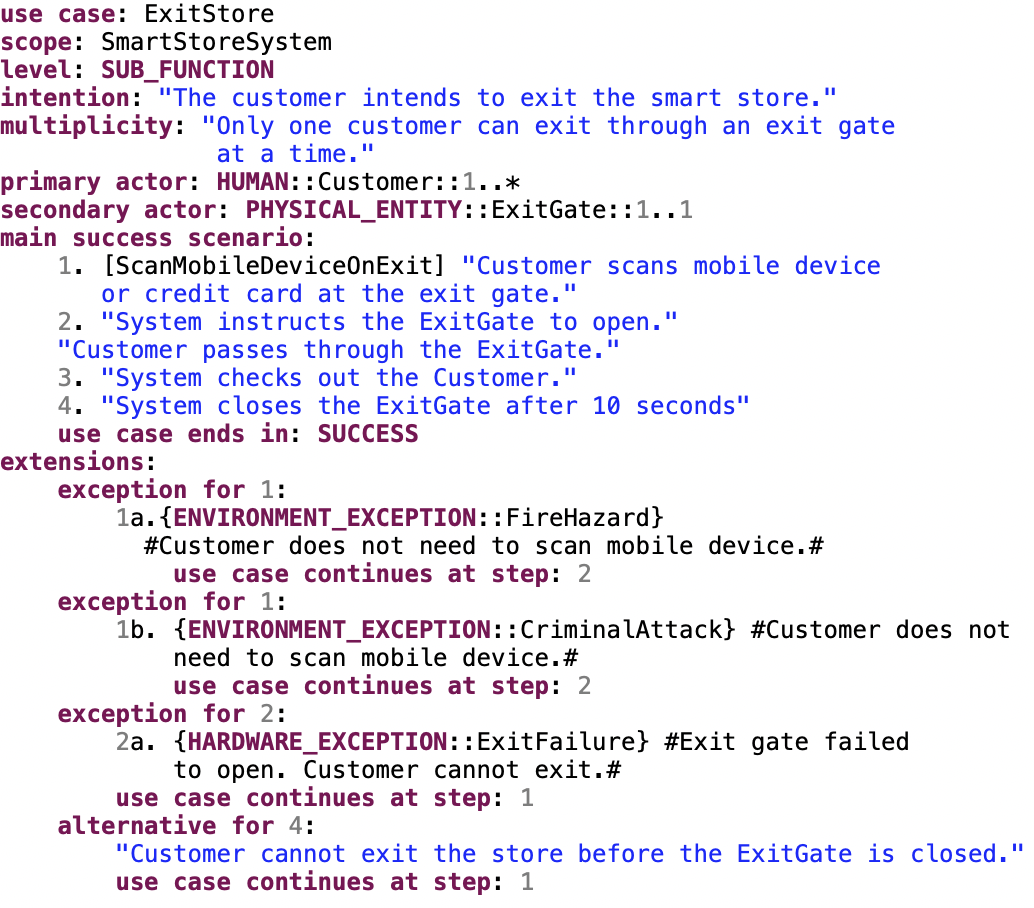}
	\caption{Smart store: Exit store use case.}
	\label{uc:exit}
\end{minipage}	
\end{figure*}	
Figure~\ref{uc:shopping} presents the summary level use case of the smart store system modelled with the UCM4IoT environment. As depicted in this use case, the primary actor, customer, has three subgoals: \texttt{Enter Store}, \texttt{Add To Cart}, and \texttt{Exit Store}. 
The \texttt{Enter Store} use case requires other actors to participate in fulfilling the customer's (\texttt{Human::Customer}) goal: \texttt{Sensor::Camera} and \texttt{PhysicalEntity::EntryGate} as secondary actors; and \texttt{PhysicalEntity::MobileDevice} and \texttt{PhysicalEntity::CrediCard} as facilitator actors. This use case invokes \texttt{Scan Mobile Device on Entry}, a sub-function level use case that includes the interaction steps associated with a customer scanning the mobile device or a credit card at the entry gate. The entry gate (\texttt{PhysicalEntity::EntryGate}) and the exit gate (\texttt{PhysicalEntity::ExitGate}) are built with readers and actuators. System reads user's device or credit card with the reader attached to these gates, and the actuator opens and closes the gates based on the system's instruction. If the store's maximum allowed capacity is reached (for instance, during a pandemic), entry is deactivated. A customer may change his mind while entering the store, and leave the store just after scanning his device at the \texttt{PhysicalEntity::EntryGate}, leading the use case to end in an \textit{abandoned} outcome.
\texttt{Add To Cart} use case starts with the customer picking an item from the shelf and ends in success after the system identifies the item (\texttt{Identify Item}) and associates the customer via image recognition (\texttt{Recognize User}). 

\begin{figure*}[!tbh]
\begin{minipage}{.5\textwidth}
  \includegraphics[width=1.2\textwidth]{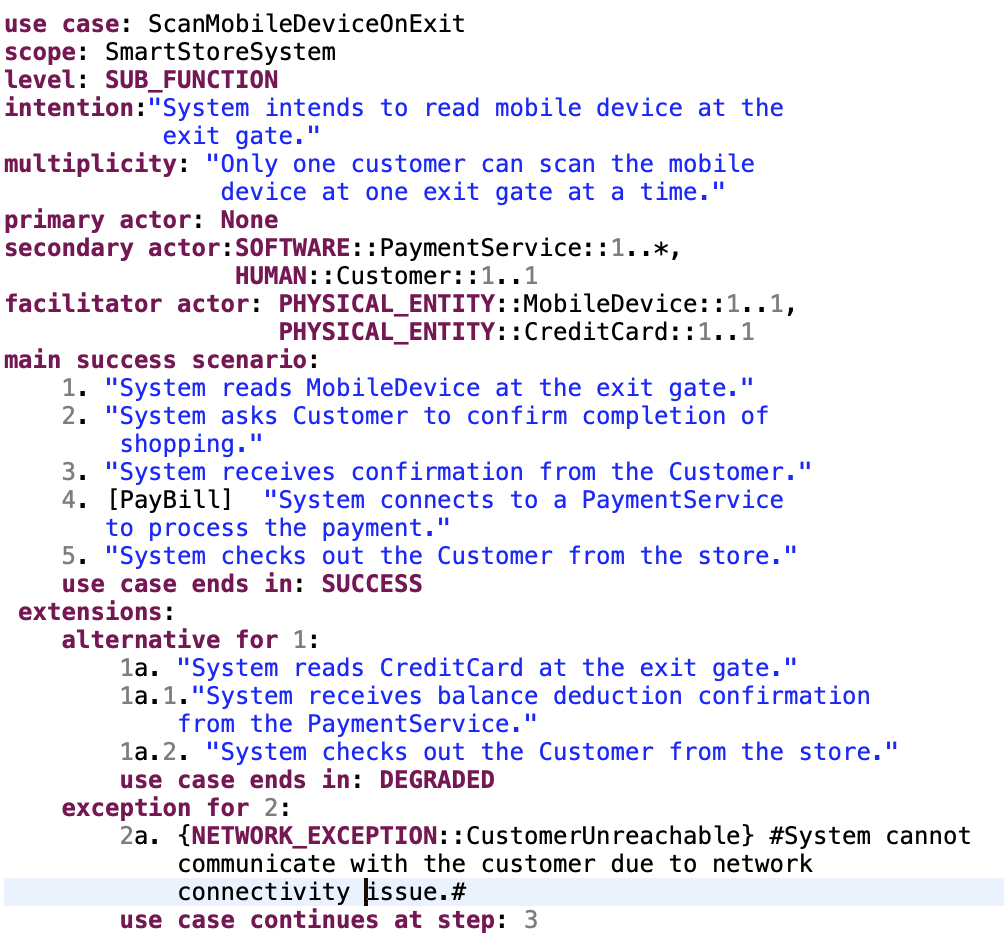}
    \caption{Smart store: Scan mobile device on exit use case.}
    \label{uc:scandevice}
\end{minipage}
\begin{minipage}{.5\textwidth}
		\centering
			\includegraphics[width=1.2
			\textwidth]{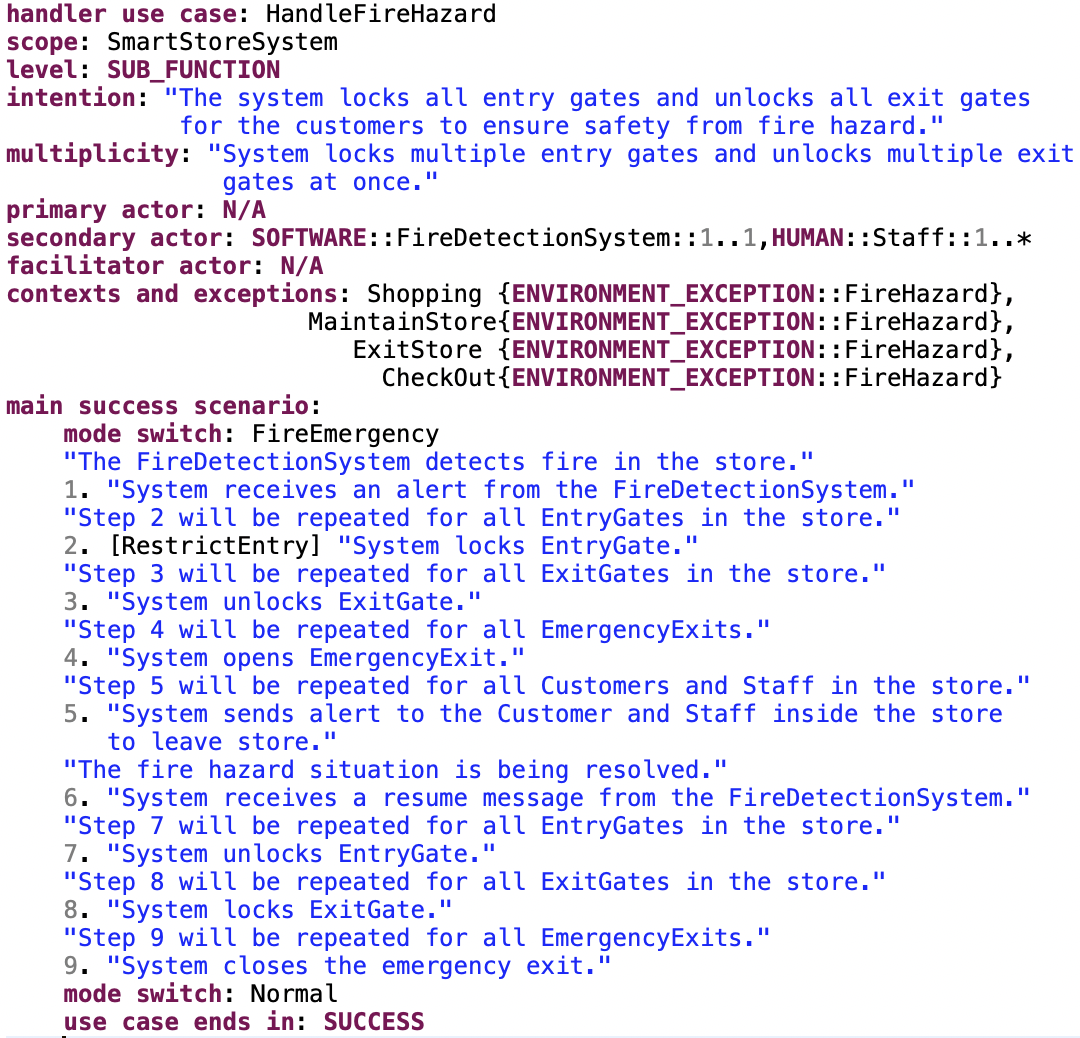}
		\caption{Smart store: Send fire alert handler.}
	\label{uc:fire}
	\end{minipage}
\end{figure*}

\texttt{Identify Item} and \texttt{Recognize User} are two core use case scenarios that make the store a ``smart store" by automatically detecting who (customer) took what (item). The secondary actors, \texttt{Sensor::PressureSensor}, \texttt{Sensor::WeightSensor}, and \texttt{Reader::TagReader}, participate in \texttt{Identify Item} to fulfill the system's goal of identifying any item on the shelves. System uses images received from \texttt{Sensor::Camera} to \texttt{Recognize User}. The interactions part of the exit scenario at the end of a shopping trip is described in the use case \texttt{Exit Store} (see Fig.~\ref{uc:exit}). The \texttt{Human::Customer} scans his/her mobile device (or credit card) at the gate invoking the interactions part of \texttt{Scan Mobile Device on Exit} (see Fig.~\ref{uc:scandevice}). 
The system processes the payment via an external \texttt{Software::Payment Service} at this time. 
Due to space constraints, only a subset of the use cases are discussed and presented here.

While developing the use case scenarios for the smart store system with UCM4IoT, we have discovered several hardware, software, network, and environment exceptions. 

\subsubsection{Exceptions and Handlers}

\textbf{Environment Exception:} 
We identified environment exceptions by considering possible abnormal scenarios occurring in the environment in which the smart store is operating that may prevent a customer or staff from fulfilling their primary goal. 

In the event of a fire hazard, the system receives input from the fire detection system (\texttt{Software::FireDetectionSystem}), an external system used to detect fire, and raises an exception (\texttt{EnvironmentException::FireHazard}, see Fig.~\ref{uc:shopping}). In this exceptional situation, no new customers are allowed to enter the store, and all remaining customers and staff need to leave the store immediately. To handle this exception, the system will temporarily lock the entry (\texttt{PhysicalEntity::EntryGate}) for new customers while the fire fighters try to mitigate the fire hazard. 

The adaptive behaviour of the system in such a scenario is defined in the \texttt{Handle Fire Hazard} handler use case (see Fig.~\ref{uc:fire}). Another environmental exception identified is any kind of criminal attack, such as robbery or a terrorist attack (\texttt{EnvironmentException::CriminalAttack}). The handling mechanism, \texttt{Alert On Attack}, is triggered when a customer or staff observes such a situation and hits one of the emergency switches (\texttt{Device::AttackAlertSwitch}) installed in the store. System notifies the nearest police station, locks the entry gates, and opens the emergency exits. 

\textbf{Hardware Exception:} With the numerous devices that are part of a smart store, hardware-related exceptions are very common for this system. 
As \texttt{Sensor::PressureSensor}, \texttt{Sensor::WeightSensor}, and \texttt{Reader::TagReader} send item data to the system at a time, if the system detects any of these three inputs are missing, it raises one or more of the following exceptions: \texttt{HardwareException::PressureUndetected}, \texttt{HardwareException::WeightUndetected}, and \texttt{HardwareException::TagUnavailable}). If the exit or entry gates stop working due to a hardware failure, \texttt{HardwareException::EntryFailure} and \texttt{HardwareException::ExitFailure} are identified. These exceptions require repair or replacement of the damaged device. In this case, a service person (\texttt{Human::ServicePerson}), is an exceptional actor (of type human user) that the system needs to handle such exceptional scenarios. 
The interactions of this exceptional actor with the system are detailed in the handler use cases, \texttt{Service Gate} and \texttt{Service Sensor}. 
The system will request the customer to use another entry gate (during check-in) or exit gate (during check-out). For item-related hardware exceptions, when the system does not receive a response from the sensor or tag reader, it sends a request along with the device location to the staff to manually scan items for the customers.

Cameras installed at the entry location may be unable to capture the necessary features of a customer (\texttt{HardwareException::CameraFailure}), causing the system to not recognize that customer. System requests customer to wait while cameras attempt to capture images (described in the \texttt{Request Customer} handler use case).

\textbf{Network Exception:} IoT systems are typically connected with various external networks. Although in the requirements elicitation phase we are assuming reliable communication between components of the smart store system under development, the system needs to communicate with external systems through a network (e.g. a customer pays via external payment services, system sends cart updates to a customer's mobile device). 
If the customer does not have network connectivity, the system will be unable to communicate with the customer to get invoice confirmation, and hence will be unable to process payment on exit. In such a scenario, a network exception, \texttt{NetworkException::CustomerUnreachable}, should be raised. 
This is handled by the system sending request to the customer to confirm payment every five seconds and waiting for the response. When the customer gets connected, he/she receives the request from the system, confirms bill payment, and exits the store. 

\textbf{Software Exception:} The smart store system needs to interact with several external software applications (e.g. payment service, fire detection system, etc.) to provide necessary system functionality. If any payment service is down, the system will not be able to process payment from customers, which ultimately disallows customers to exit the store with the items. System checks payment status at the exit gate when someone scans the mobile device, and payment failure will cause the system to not instruct the exit gate to open by default (\texttt{SoftwareException::PaymentServiceDown}). To handle this scenario, a special system feature is required such that the system updates the payment status to \emph{on hold} and lets the customer leave the smart store. The handler \texttt{Hold Payment} 
describes the system's adaptive behaviour. 

\texttt{SoftwareException::PhotoUnmatched} is another example of a software exception raised in the \texttt{Recognize User} use case.

\begin{figure}[tbh!]
\centering
 \includegraphics[width=1.1\textwidth]{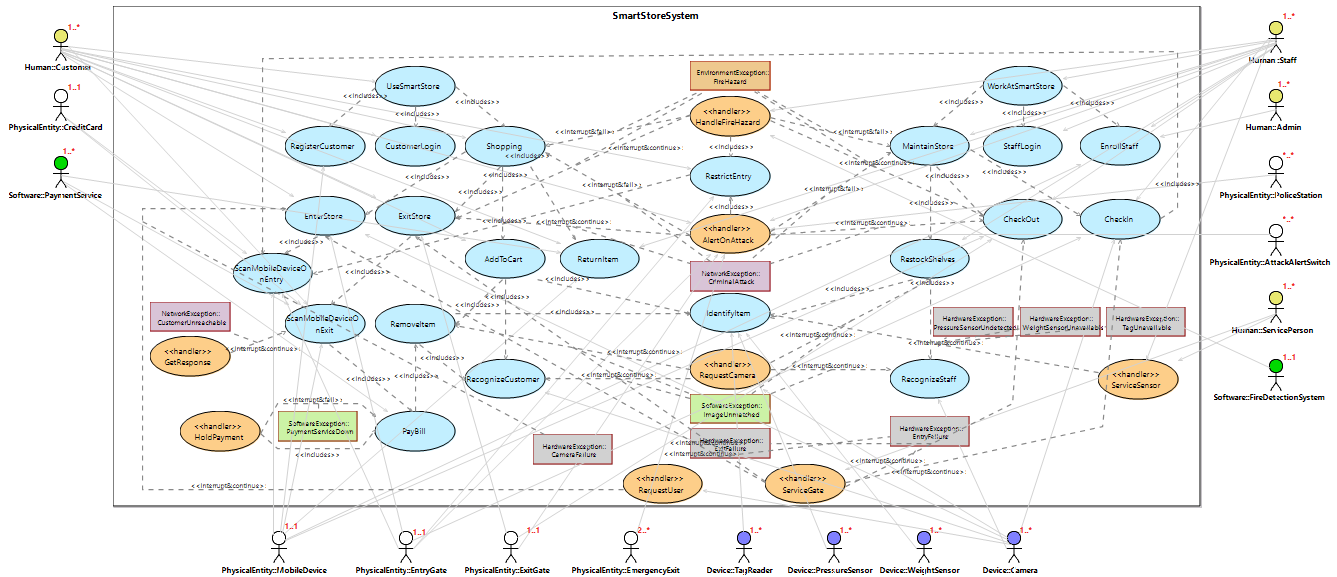}
     \caption{Smart store system: Extended use case diagram.} 
     \label{fig:ssucd}
\end{figure}

 Figure \ref{fig:ssucd} presents the use case diagram of the smart store system. To model the structural view of the system, an IoT ARM domain model was also developed. The UCM4IoT model in conjunction with this domain model form the basis for the subsequent analysis and design. 

\subsubsection{Services and Modes}

Services are defined after completing smart store system's requirements elicitation with UCM4IoT. Services were identified to meet the requirements of this system. The services with the associated goals are listed below.

\begin{tiny}
\begin{enumerate}
     \item Cart processor: \texttt{Add To Cart}, \texttt{Return Item}, \texttt{Remove Item}
    \item Bill payer: \texttt{Pay Bill}, \texttt{Hold Payment}
    \item User recognizer: \texttt{Recognize Customer}, \texttt{Recognize Staff}, \texttt{Check In}, \texttt{Request User}, \texttt{Request Camera}
    \item Entry operator: \texttt{Scan Mobile Device On Entry}, \texttt{Check In}, \texttt{Service Gate}
     \item Exit operator: \texttt{Scan Mobile Device On Exit},  \texttt{Check Out}, \texttt{Service Gate}, \texttt{Get Response}
     \item Entry restriction manager:\texttt{Restrict Entry}
    \item Access manager: \texttt{Staff}, \texttt{Staff Login}, \texttt{Customer Login}, \texttt{Register Customer}
    \item Inventory manager: \texttt{Restock Shelves}
     \item Police station notifier: \texttt{Alert On Attack}
    \item Fire hazard manager: \texttt{Handle Fire Hazard}
\end{enumerate}
\end{tiny}

\textit{Cart processor} uses \textit{User recognizer} and \textit{Item identifier} services to process customers' cart. It matches customer with their picked or returned items and updates cart accordingly. It then sends update to the \textit{Inventory manager} service. \textit{Bill payer} service receives updated cart information from the \textit{Cart processor} and connects to the payment services to process the payment.

\textit{Item identifier} service identifies items in the shelf with sensor data. \textit{User recognizer} service recognize customers and staff with the information sent by cameras. \textit{Entry operator} service controls all entry, and \textit{Exit operator} controls exit gates. \textit{Access manager} service manages access to the smart store system. 
  
\textit{Fire hazard manager} service operates in case of a fire detected by an external \texttt{FireDetectionSystem}. \texttt{HandleFireHazard} handler invokes a change in behaviour of the \textit{Bill payer} service. \textit{Police station notifier} service triggered by actor \texttt{AttackAlertSwitch} communicates with the nearest police station. \texttt{AlertOnAttack} handler changes the behaviour of the \textit{Cart processor} and \textit{Exit operator} services. In case of a fire hazard or criminal attack, all exit gates unlock automatically and remain open, temporarily suspending the requirement for scanning customer devices on exit. All emergency gates also remain open in these two emergency cases. However, entry gates stay locked (\textit{Entry restriction manager} service) in these situations to restrict any new customer to enter the store.

\begin{figure}[!tbh]
\centering
    \includegraphics[width=0.5\textwidth]{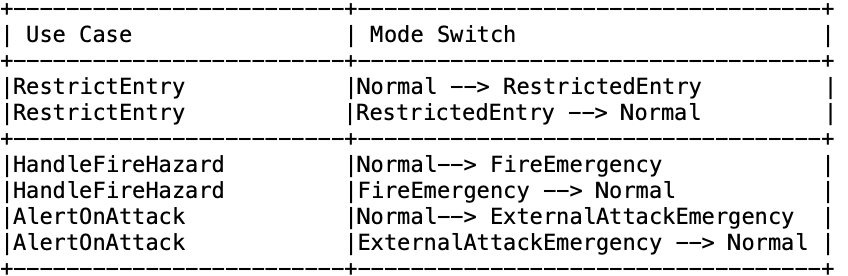}
    \caption{Smart store system: Generated mode switch table.}
    \label{fig:modeTable}
\end{figure}

\begin{table}[!tbh]
\fontsize{7}{8}\selectfont
\centering
  \caption{Smart Store System: Mode Summary Table} \label{ss-modesummary}
\begin{tabular}{|p{3.3cm}|p{2cm}|p{6cm}|}
\hline
\textbf{Modes} & \textbf{Type} & \textbf{Available Services}\\
\hline
Normal & Normal & All services excluding \textit{Entry Restriction Manager}, \textit{Fire Hazard Manager}, and \textit{Police Station Notifier} \\
\hline
Restricted Entry & Restricted & \textit{Entry Restriction Manager} and all other services excluding \textit{Entry operator}, \textit{Fire Hazard Manager} and \textit{Police Station Notifier} \\
\hline
Fire Emergency & Emergency & \textit{Entry Restriction Manager}, \textit{Fire Hazard Manager}\\
\hline
External Attack Emergency & Emergency & \textit{Entry Restriction Manager}, \textit{Police Station Notifier}\\
\hline
\end{tabular}
\end{table}

Different services are activated in different modes. The following section discusses smart store system's operational modes and the services.

\textbf{Normal Mode}: All services operate in \texttt{Normal} mode except \texttt{Entry restriction manager}, \texttt{Fire hazard manager}, and \texttt{Police station notifier}. Smart store system has only one \texttt{Normal} mode for regular operations. A mode switch from \texttt{Normal} to some other mode occurs in an alternate step (e.g., switching from \texttt{Normal} to \texttt{RestrictedEntry)}), and in other exceptional situations (\texttt{EnvironmentException::FireHazard}). All mode switches of the smart store system is shown in Fig.~\ref{fig:modeTable}. 

\textbf{Emergency Mode}: There are two emergency modes in the smart store system. The environment exceptions (\texttt{EnvironmentException::FireHazard} and \texttt{EnvironmentException::CriminalAttack}) impact the services offered by the system, hence triggering a mode change to the \texttt{Fire Emergency} and \texttt{External Attack Emergency} mode respectively (see Fig.~\ref{fig:modeTable}). Handling mechanisms are invoked to recover from the emergency situation (\texttt{Handler Fire Hazard}, \texttt{Alert On Attack}). On recovery, the system resumes operation in the \texttt{Normal} mode. 

\textbf{Restricted Mode}:
In a situation where customer entry needs to be restricted, the system can still offer the checkout service but no new customer can be allowed into the store. In such a case, the system moves to a \texttt{Restricted Entry} mode offering restricted services. This situation may arise when there is the store needs to operate with a limited capacity, such as in a pandemic. Table ~\ref{ss-modesummary} shows the services available in a particular mode for smart store system.

\subsection{Exception and Handler Summary Generation and Analysis}

Figure \ref{fig:tool-validation} presents the generated exception summary table for the smart store system. Based on this information, it is possible to quickly identify the possible paths (sequence of use cases) that result in an exception being thrown. Modellers can identify exceptions that are raised in many possible sequences. They can then determine how critical that use case, exception, and any related handlers are to the rest of the system.

As an example, the \texttt{TagUnavailable}, \texttt{PressureSensorUndetected}, and \texttt{WeightSensorUnavailable} exceptions can occur in three possible sequences of use cases each: \textsl{UseSmartStore$\rightarrow$ Shopping $\rightarrow$ AddToCart $\rightarrow$ IdentifyItem}, \textsl{UseSmartStore$\rightarrow$Shopping $\rightarrow$ AddToCart $\rightarrow$ RemoveItem $\rightarrow$ IdentifyItem}, and \textsl{UseSmartStore$\rightarrow$Shopping $\rightarrow$ ExitStore $\rightarrow$ ScanMobileDeviceOnExit $\rightarrow$ PayBill $\rightarrow$ RemoveItem $\rightarrow$ IdentifyItem}. Likewise, the \texttt{Service Sensor} handler handles each occurrence of any of those three exceptions. As each exception can be raised in three possible sequences each, the handler can be invoked in a total of nine different paths. Hence, we can deduce the significance of this handler in this system and attach a higher weight or priority to it.

The mode summary table (Table~\ref{ss-modesummary}) shows that if there is a fire emergency, customers and staff will not be able to avail any other services apart from exiting the store. The generated mode switch information, shown in Figure~\ref{fig:modeTable}, explicitly shows the use cases in which the mode of operation switches back to normal. Hence, the success of these goals can be inferred to be critical to the normal operation of the system, requiring designers to ensure that the level of quality-of-service provided by this service is sufficient to continue with normal operation.

%% file: casestudy_sfa.tex
\section{Case Study 2: Smart Fire Alarm System}
\label{sec:casestudy-sfa}

In this section, UCM4IoT is demonstrated with the smart fire alarm system case study.

\subsection{Smart Fire Alarm System Overview}

Over the last decade, IoT devices have revolutionized many industries, from retail and healthcare to home appliances. However, most residential and commercial buildings still use traditional fire detection systems. Smart fire alarm systems are becoming more common as they are safer than conventional systems as they automatically alert the fire department and the user when a fire is detected. A smart fire alarm system is a system that utilizes internet connectivity to automatically alert the fire department and user when a fire is detected. These smart systems are safer than traditional systems as quick notification increases fire response time and reduces deaths.

The system can also communicate with other systems, such as a sprinkler system, and automatically trigger it. 
Implementing this system requires various sensors such as heat, smoke, and carbon monoxide. It also requires an initial setup. Users are required to input their contact information and the contact information of their local fire department. 

If any of the sensors are triggered, the system automatically alerts the user(s) and the fire department. The heat sensor has a connection with the sprinkler system, and if it detects a high level of heat, it will communicate with the sprinkler system and request it to turn on the sprinkler nearest to the detected heat source. After the initial setup, the system does not require any human intervention unless a hardware failure requires maintenance.

Hardware and software requirements for the smart fire alarm system are initially derived based on domain knowledge and informal requirements. Next, the UCM4IoT language is used to specify the system requirements and further elicit any applicable hardware, software, network, and environment exceptions.

\subsection{Requirements Development with UCM4IoT}

\subsubsection
{Actors and Goals} The Smart Fire Alarm System has four primary actors, User (\texttt{Human::User}), HeatSensor (\texttt{SENSOR::HeatSensor}), SmokeSensor (\texttt{SENSOR::SmokeSensor}) and CarbonMonoxideSensor (\texttt{SENSOR::CarbonMonoxideSensor}).

\begin{figure*}[!tbh]
\begin{minipage}{.5\textwidth}
  \includegraphics[width=.9\textwidth]{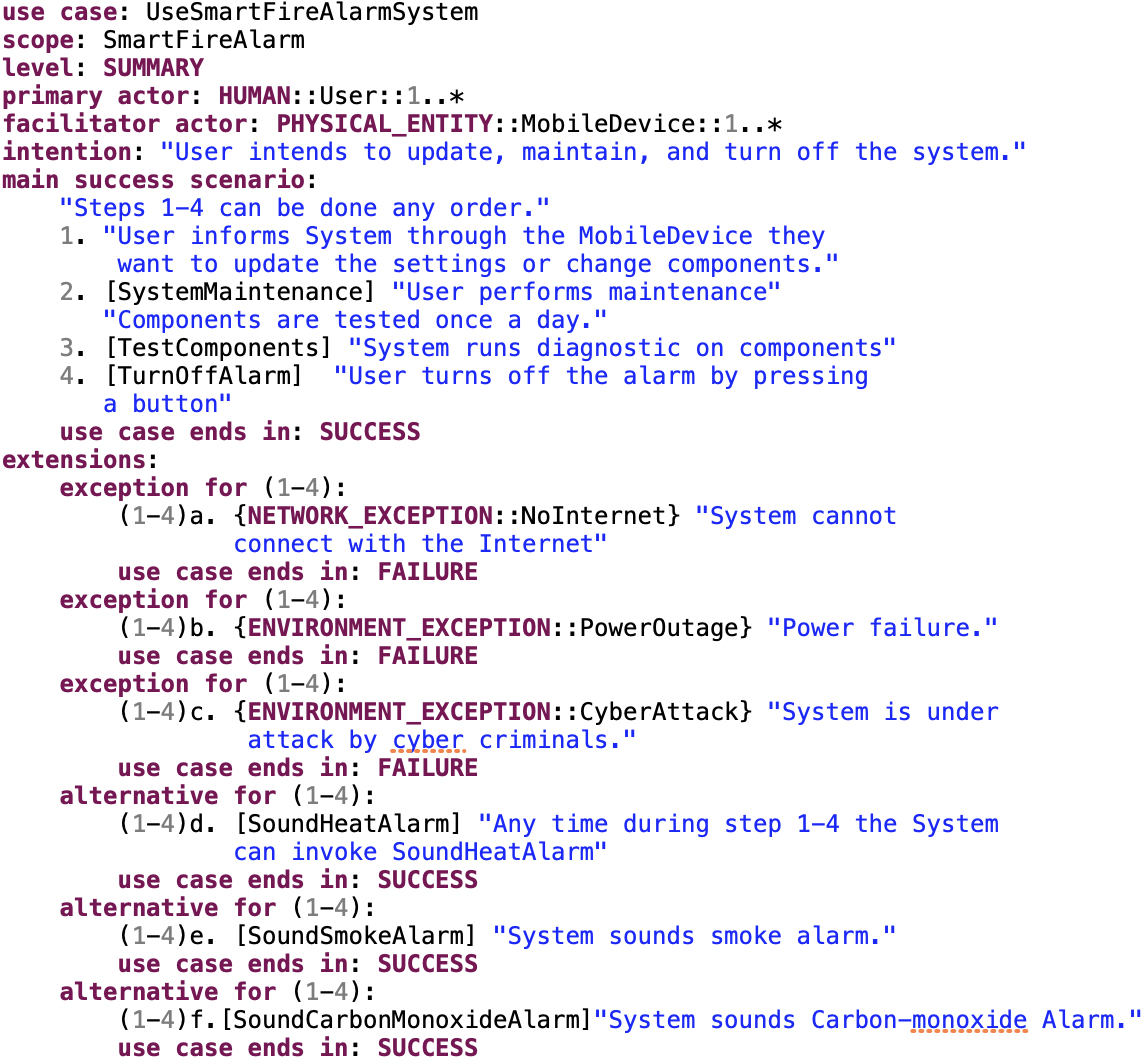}
    \caption{Smart fire alarm: Use smart fire alarm system use case.}
    \label{uc:sfa}
\end{minipage}
\begin{minipage}{.5\textwidth}
		\centering
			\includegraphics[width=1\textwidth]{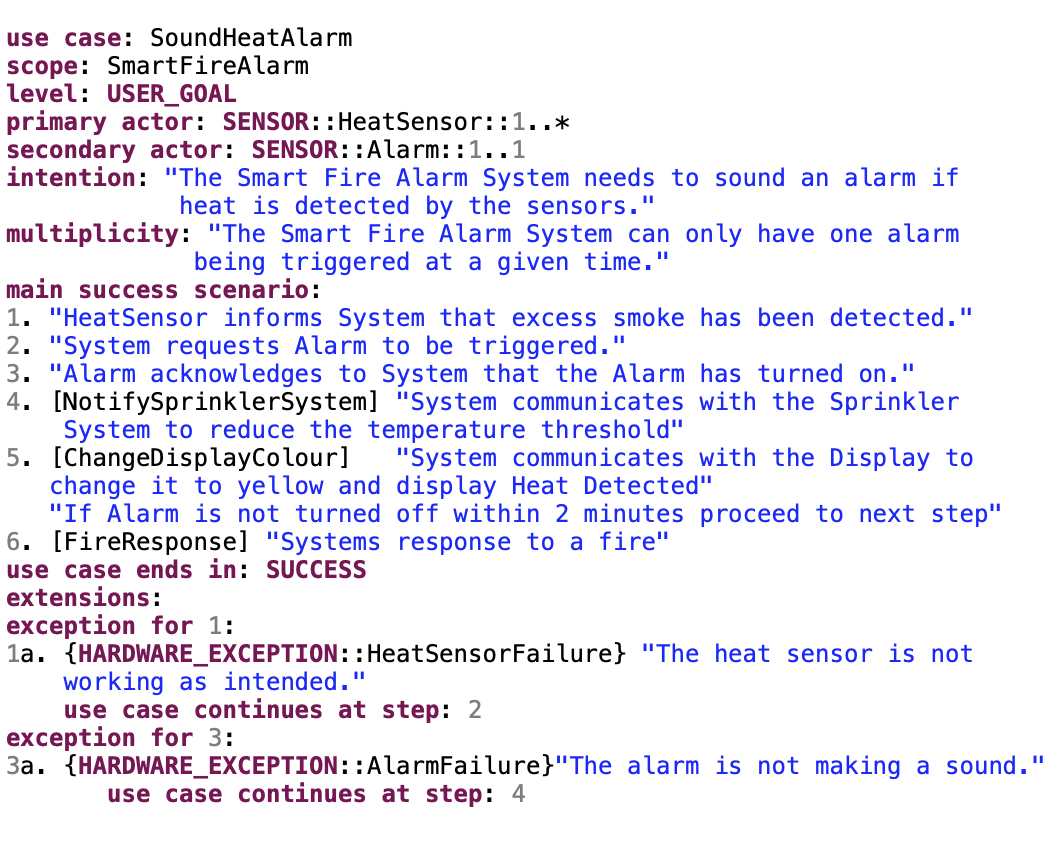}
		\caption{Smart fire alarm: Sound heat alarm use case.}
	\label{uc:heat}
	\end{minipage}
\end{figure*}

One of the summary level use cases for the smart fire alarm system that was written using the UCMIoT tool is presented in Fig.~\ref{uc:sfa}. The primary actor, the user, has three sub-goals: System Maintenance, Test Components, and Turn Off Alarm. The System Maintenance use case does not require any other actor and does not need to invoke other use cases to fulfill the user goal. In this scenario, the user first authenticates with the system by providing a password and then updates the system settings. Next, the system refreshes the software to synchronize any new changes. If the authentication fails, the use case ends in failure. 

The user's goal with the system is to perform system maintenance and turn off the fire alarm. To achieve this goal, the user enters a password to authenticate with the system and then replaces the faulty sensors/battery. The user then requests the system to refresh the software to ensure the new component is synced correctly. The user's second goal is to turn off the alarm; this is done by pressing the turn off button on the alarm, which makes the system request the alarm to be turned off. The user's goal will be successful if they are able to perform system maintenance by replacing a component and also if they are able to turn off the alarm manually. The user's goal will have a failed outcome if the user cannot authenticate with the system or if the turn-off button is not functioning. Participating actors have been categorized as per the UCM4IoT actor types.

The goal of the heat sensor is to sense heat and inform the system so it can trigger the alarm. Fig.~\ref{uc:heat} illustrates the \texttt{Sound Heat Alarm} use case. The sensor will then invoke the fire response use case. Similarly, the smoke sensor's goal is to sense smoke levels, and the carbon monoxide sensor's goal is to sense carbon monoxide levels. The sensor's primary goal will be successful if the alarm is triggered and the fire response use case is invoked. The alarm is triggered when abnormal amounts of heat, smoke, or carbon monoxide are detected. These trigger the \texttt{Alert Fire Department} use case. 
The fire response use case is invoked if heat, smoke or carbon monoxide are sensed for more than 2 minutes (the timeout may vary for different systems). The sensor's goal will have a failed outcome if there is a hardware failure with the sensors. 

The secondary actors, \texttt{Sensor::HeatSensor}, \texttt{Sensor::SmokeSensor}, and  \texttt{Sensor::CarbonMonoxideSensor}, \texttt{PhysicalEntity::Battery}, are required to complete the Test Components goal. The use case starts with the system checking the battery level and then notifying the User if it is below a certain threshold through their mobile device. Next, the system runs diagnostic tests on all the various sensors, and if it finds a failing sensor, it will notify the User through the mobile device. However, if the Diagnostic test shows all the sensors are working correctly, the use case will end in success.

The \texttt{Fire Response} use case will invoke all the other required use cases to handle the fire. The first use case invoked is \texttt{Notify Sprinkler System}, in which the system communicates with the Sprinkler System to reduce the temperature threshold.  The second use case invoked is \texttt{Alert Fire Department}, in which the system notifies the fire department about the fire.  Then, it will invoke the \texttt{Alert User} use case in which the system alerts the User about the fire. 
It will also invoke the \texttt{Change Display Colour} use case in which the system communicates with the Display to change the colour to red.  Finally, it will invoke the \texttt{Open Emergency Door} use case in which the system requests the emergency door to unlock. 

Numerous software, hardware, network, and environment exceptions were discovered during the creation of use cases for the smart fire alarm system using the UCMIoT language.

\begin{figure*}[tbh!]
\begin{minipage}{.5\textwidth}
\includegraphics[width=1\textwidth]{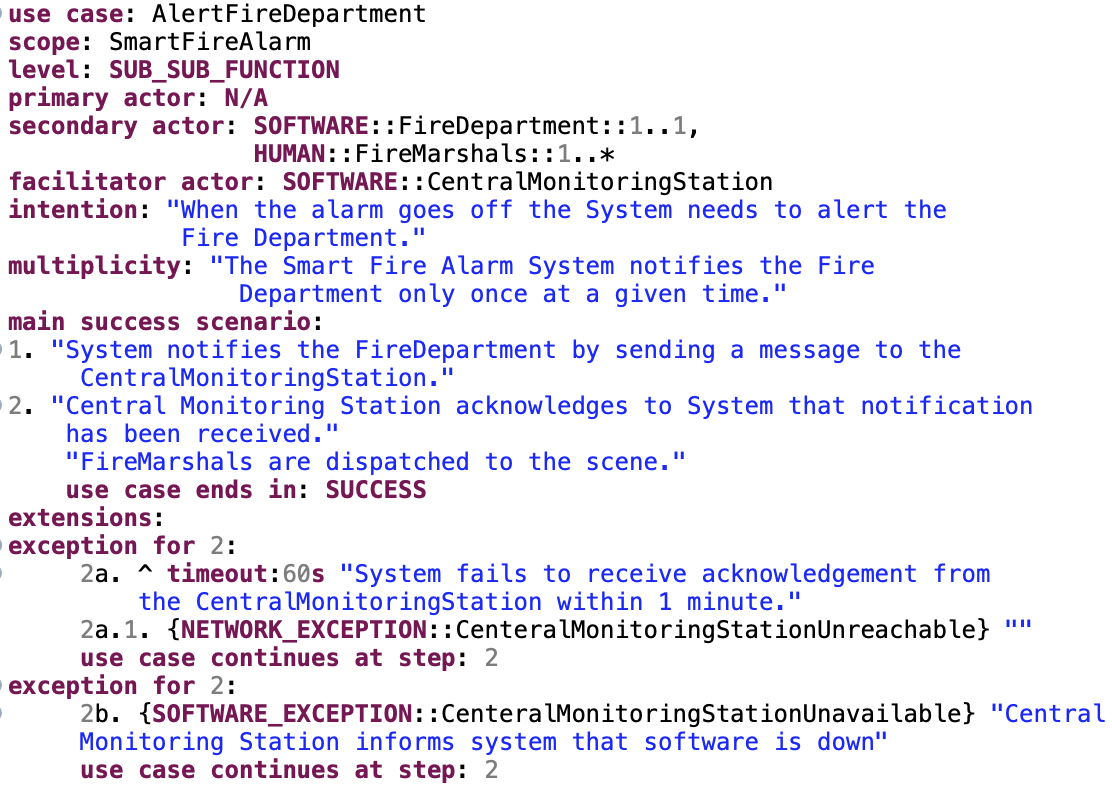}
    \caption{Smart fire alarm: Alert Fire Department use case.}
    \label{uc:alert}
    \end{minipage}
\begin{minipage}{.5\textwidth}
 \includegraphics[width=1\textwidth]{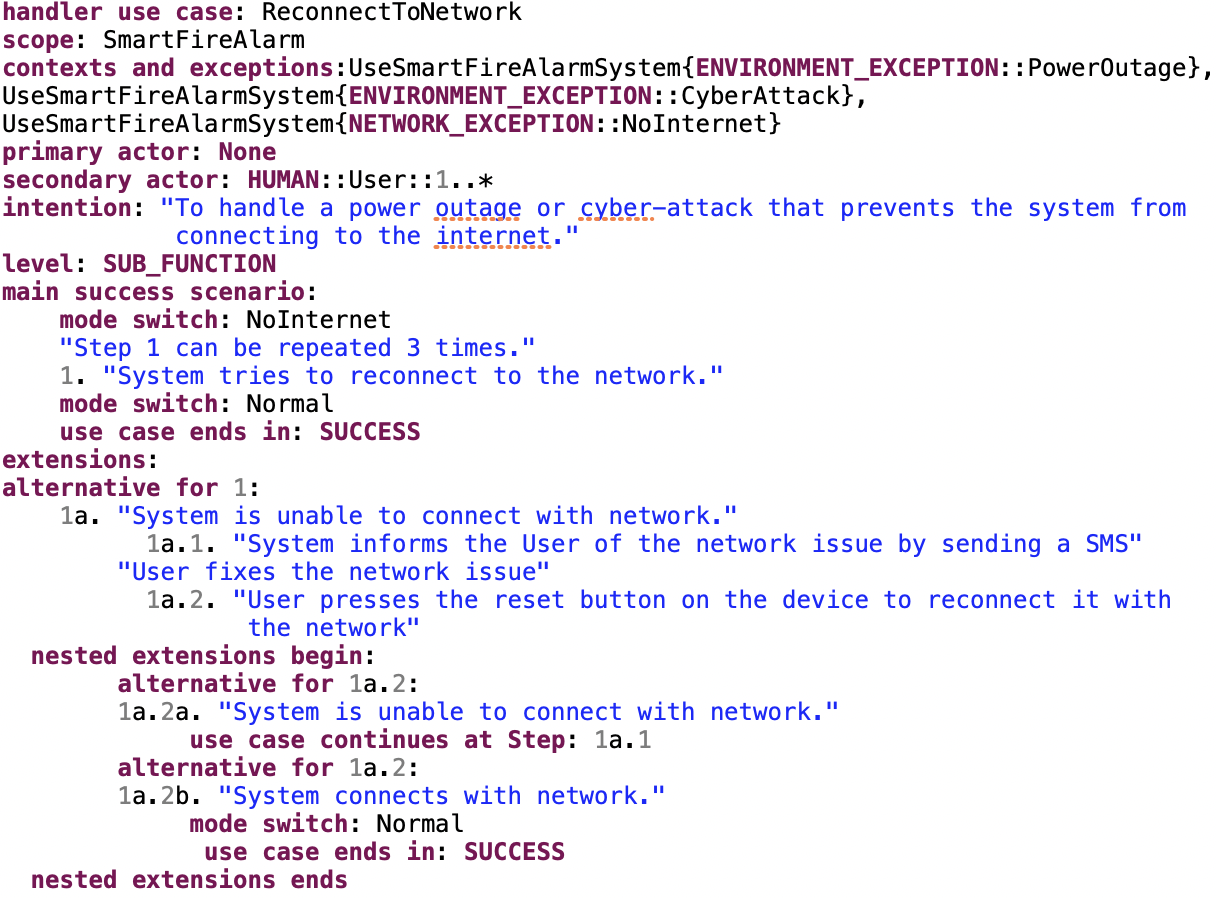}
    \caption{Smart fire alarm: Reconnect To Network handler use case.}
    \label{uc:recon}
\end{minipage}
\end{figure*}

\subsubsection{Exceptions and Handlers} \textbf{Environment Exception:}  
In the event that a power outage occurs, the system will raise an exception \texttt{EnvironmentException::PowerOutage}, see Fig. 4). When this exception occurs, the network connectivity will be lost, so the system will be unable to alert the User and fire department if a fire is detected. The system will also not be able to communicate with the sprinkler system. To handle this exception, the system first tries to reconnect to the network a maximum of three times. If the issue is still not resolved, the system will switch to a \texttt{NoInternet} mode and alert the User about the network issue by sending an SMS. Once the issue is resolved, it will switch back to the normal mode. The adaptive behaviour of the system in such a scenario is defined in the \texttt{ReconnectToNetwork} handler use case. Another environmental exception that might occur is a cyberattack \texttt{EnvironmentException::CyberAttack}. The cyberattack can occur at any time, and it will compromise the system by taking down the internet connection. This exception is handled the same way as the \texttt{PowerOutage} using the \texttt{ReconnetToNetwork} handler. Fig.~\ref{uc:recon} shows the use case to handle these exceptions.

\textbf{Hardware Exception:} The smart fire alarm system utilizes various sensors and hardware components to operate, and as a result, hardware exceptions can occur frequently. To detect the fire or carbon monoxide, the \texttt{Sensor::SmokeSensor}, \texttt{Sensor::HeatSensor}, and \texttt{Sensor::CarbonMonoxideSensor} are used. If a system test determines that any of these sensors are not working, it will raise one of these exceptions \texttt{HardwareException::SmokeSensorFailure}, \texttt{HardwareException::HeatSensorFailure}, or \texttt{HardwareException::CarbonMonoxideSensorFailure}. The physical button to turn off the Alarm and reset the sensors might not work, raising the \texttt{HardwareException::ButtonFailure}. 

One of the essential hardware components is the Alarm; if one or more sensors are triggered but the Alarm is not sounding, then the \texttt{HardwareException::AlarmFailure} is called. 
The Display might have a hardware failure \texttt{HardwareException::DisplayColorChange} or \texttt{HardwareException::DisplayColorRevert}, resulting in disabled individuals not being alerted about the fire. If the automatic emergency door does not work, \texttt{HardwareException::DoorNotWorking} is raised. All of these hardware exceptions require repairing the damaged sensor or physical entity. The system will need the User \texttt{Human::User} to handle such exceptional scenarios. The interaction steps of this exceptional actor with the system are defined in the System Maintenance handler use cases.

\textbf{Network Exception:}  
If the system gets disconnected from the network, a network exception, \texttt{NetworkException::NoInternet}, will be raised. This exception is handled by the \texttt{ReconnectToNetwork} handler, where the system first tries to reconnect to the network three times. If it still cannot connect, the system will inform the User of the network issue by sending an SMS. After the User resolves the issue, they will press the reset button to let the system know. Then the system will try to reconnect; if it still does not work, the system will keep looping the above steps. Network exceptions can also be raised if the network issue comes from the external system. For example, if the sprinkler system has connection issues, the system will raise a \texttt{NetworkException::SprinklerSystemUnavailable}. Similarly, if the Central Monitoring Station or the User does not have a network connection, the following exceptions will be raised: \texttt{NetworkException::CenteralMonitoringStationUnavailable}, \texttt{NetworkException::UserUnavailable}. All three of these exceptions are also handled by the \texttt{ReconnectToNetwork handler} in a similar manner. 

\textbf{Software Exception:} The smart fire alarm system communicates with various external software systems to provide users with essential functionality.  For example, the system needs to communicate with the Central Monitoring Station software system to provide the \texttt{AlertFireDeparment} functionality. If this system is down, the \texttt{SoftwareException::CenteralMonitoringStationUnavailable} exception will be raised. Another example of a software exception is \texttt{SoftwareException::SprinklerSystemUnavailable}, where the system cannot provide the sprinkler notification functionality and send messages to the sprinkler system. 

\begin{figure}[!tbh]
    \includegraphics[width=1\textwidth]{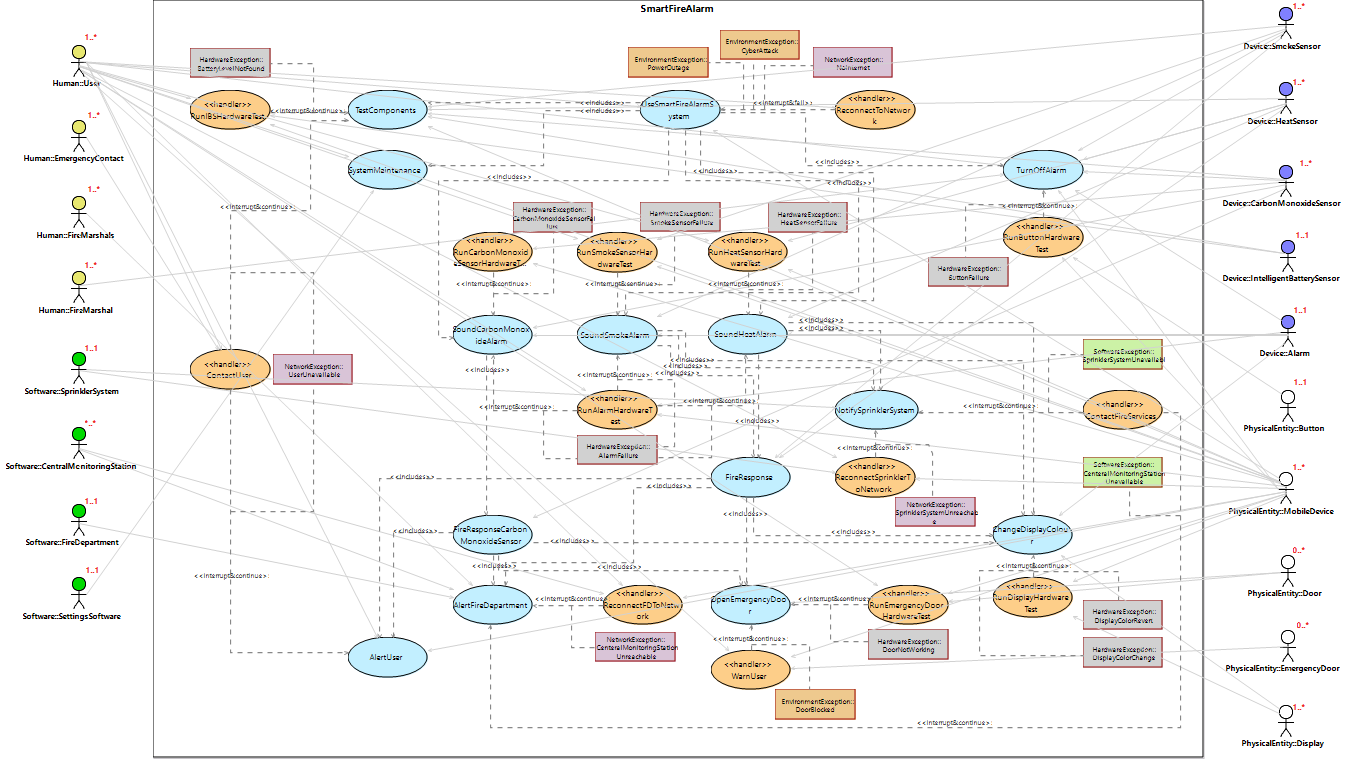}
    \caption{Smart fire alarm system: Use case diagram.}
    \label{ucd:sfa}
\end{figure}

The use case diagram showcasing the actors, use cases, exceptions, and handlers derived from the UCM4IoT model is presented in Fig.~\ref{ucd:sfa}. The UCM4IoT model forms the basis for the subsequent analysis.

\subsubsection{Services and Modes}
We discovered the following services are required to satisfy the goals of the smart fire alarm system.

\textit{Heat detector}, \textit{Smoke detector}, and \textit{Carbon-monoxide detector} services detect and sound alarm for heat, smoke, and carbon-monoxide respectively. These services invoke the \textit{Notifier} service to notify sprinkler system and alert fire department in case of a fire hazard. \textit{Notifier} service also alerts and warns the system user.

Commercial users of this smart fire alarm system use \textit{Display colour changer} service along with other services where the system display colour changed from green to red in case the system detects fire, smoke, or carbon-monoxide. The system needs regular maintenance, that is performed by the \textit{System maintainer} service. 

System needs to test its hardware components regularly to ensure maximum safety. \textit{Component tester} service tests all hardware components of the system. In case of exceptions, system also need to run tests on these components. Thus, all the hardware and network exceptions are handled by this service. Users of this system can manually turn off the alarm by using the \textit{Users control} service. \textit{Emergency doors control} service controls the doors opening after receiving notification from the \textit{Notifier} service. The following list shows services and the associated goals.

\begin{enumerate} \tiny{
    \item Heat detector: \texttt{Sound Heat Alarm}
    \item Smoke detector: \texttt{Sound Smoke Alarm}
    \item Carbon-monoxide detector: \texttt{Sound Carbon Monoxide Alarm}
    \item Notifier: \texttt{Notify Sprinkler System}, \texttt{Alert Fire Department}, \texttt{Alert User}, \texttt{Reconnect To Network}, \texttt{Warn User}
    \item Display colour changer: \texttt{Change Display Colour}
    \item System maintainer: \texttt{System Maintenance}
    \item Component tester: \texttt{Test Components}, \texttt{Run Smoke Sensor Hardware Test}, \texttt{Run Heat Sensor Hardware Test}, \texttt{Run Alarm Hardware Test},\texttt{Run Emergency Door Hardware Test}, \texttt{Run Carbon Monoxide Sensor Hardware Test}, \texttt{Run Button Hardware Test}, \texttt{Run Display Hardware Test}, \texttt{Contact User}, \texttt{Reconnect Sprinkler to Network}, \texttt{Reconnect FD to Network}, \texttt{Run IBS Hardware Test}
    \item Users control: \texttt{Turn Off Alarm}
    \item Emergency doors control: \texttt{Open Emergency Door}}
\end{enumerate}

The smart fire alarm system has two operational modes: one normal mode for regular operations and two restricted modes.

\textbf{Normal Mode}: All services are available in the\texttt{Normal} mode except when there is an issue with the network that does not let the system reach the central monitoring system in case of power outage, cyberattack, and lack of internet connection. 

\textbf{Degraded Mode}: When system cannot connect to the central station due to network exception, it switches to a degraded mode, \texttt{No Alert}. The system goes to the \texttt{No Internet} mode in case of \texttt{ENVIRONMENT\_EXCEPTION::PowerOutage}, \texttt{ENVIRONMENT\_EXCEPTION::CyberAttack}, and \texttt{NETWORK\_EXCEPTION::NoInternet}. In \texttt{No Internet} mode, all services except \textit{Component tester} and \textit{System maintainer} service will remain operational.

\begin{table}[!tbh]
\fontsize{7}{8}\selectfont
\centering
\caption{Smart fire alarm system: Mode summary table.} \label{modeSummary-sfa}
\begin{tabular}{|p{2.5cm}|p{2cm}|p{7cm}|}
\hline
\textbf{Modes} & \textbf{Type} & \textbf{All Services}\\
\hline
Normal & Normal &  All Services\\ \hline
No Alert & Degraded &  All services excluding \textit{Contact user}\\ \hline
No Internet & Degraded & All services excluding \textit{Component tester} and \textit{System maintainer}\\ \hline
\end{tabular}
\end{table}

\subsection{Exception and Handler Summary Generation and Analysis}

\begin{landscape}
\begin{figure}[tbh!]
    \centering
    \includegraphics[width=1.5\textwidth]{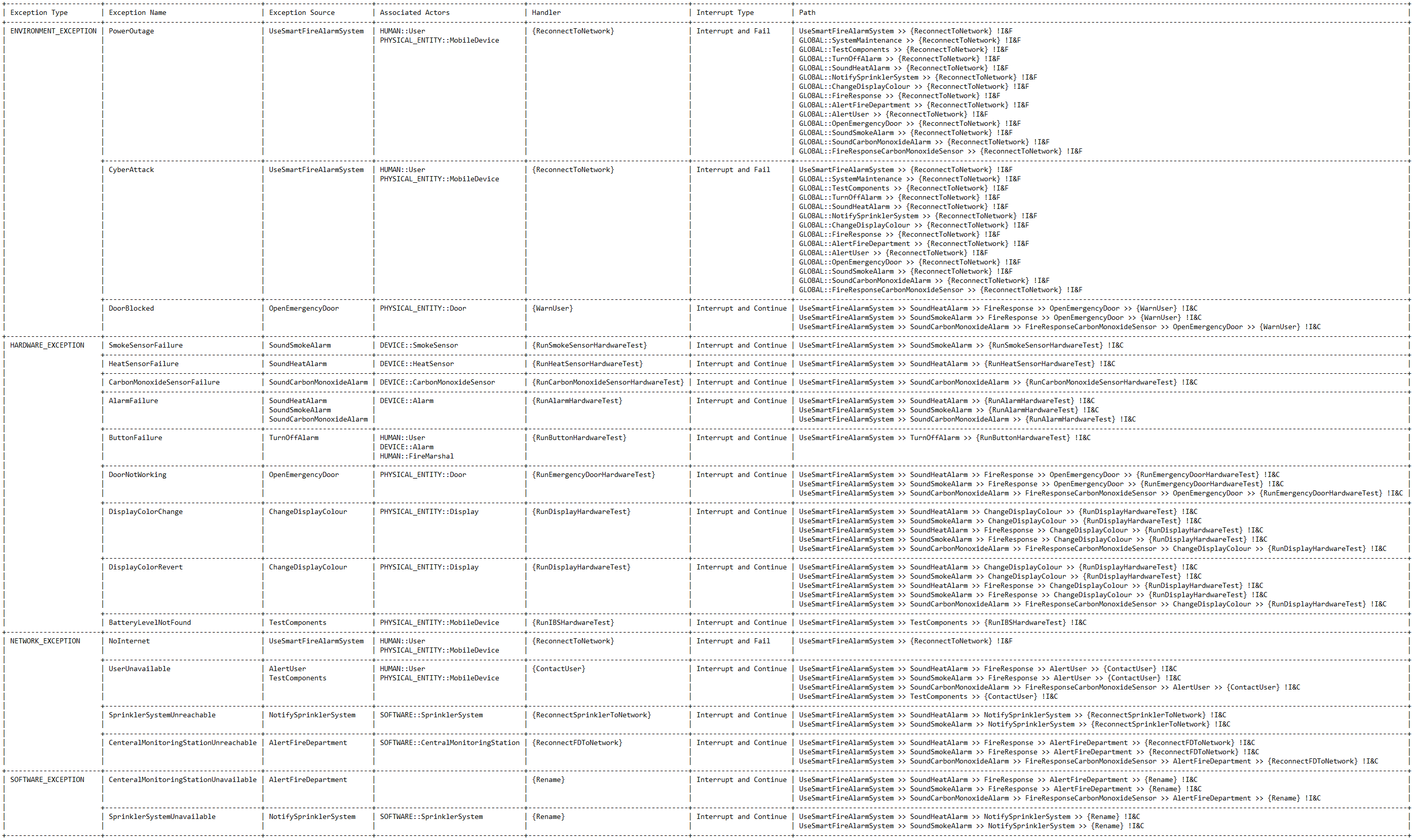}
    \caption{Smart fire alarm system: Generated exception summary.}
    \label{fig:sfa-exceptionTable}
\end{figure}
\end{landscape}

\begin{figure}[tbh!]
    \centering
    \includegraphics[width=1\textwidth]{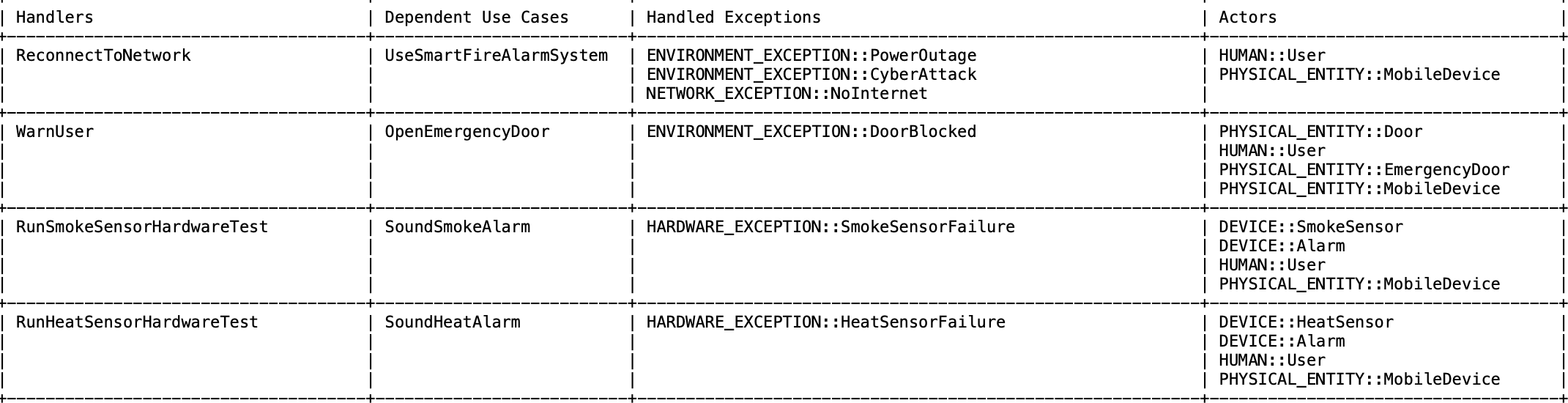}
    \caption{Smart fire alarm: Generated handler summary (slice).}
    \label{fig:sfa-handlerTable}
\end{figure}

Figure~\ref{fig:sfa-exceptionTable} presents the generated exception summary table and Fig.~\ref{fig:sfa-handlerTable} presents a slice of the generated handler summary table for the smart fire alarm system. Based on the paths generated in the summary information, we can determine the criticality of the use cases, exceptions and handlers.

As an example, the exception summary table can help in deducing that the \texttt{Change Display Color} use case has a high significance since it is included in 10 possible sequences and may raise hardware exceptions. Similarly \texttt{AlertUser} also appears in 4 possible sequences and may raise network exceptions. These use cases and associated handlers should be assigned a higher priority and/or dependability level.

By analyzing the \texttt{ReconnetToNetwork} handler, it can be seen that it is a crucial handler as it handles the \texttt{PowerOutage}, \texttt{CyberAttack}, and \texttt{NoInternet} exceptions (as seen in Fig.~\ref{fig:sfa-handlerTable}). All three of these exceptions need to be addressed immediately, since a lack of network connectivity hampers critical system functionality.

\begin{figure}[!tbh]
\centering
    \includegraphics[width=.6\textwidth]{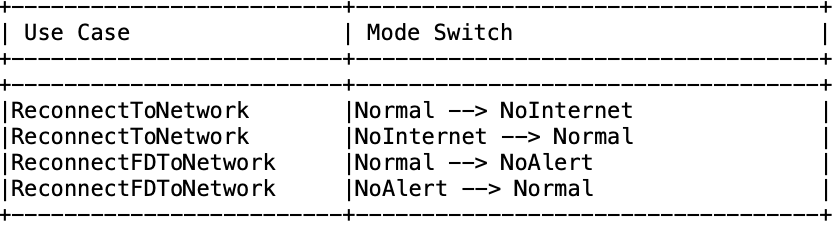}
    \caption{Smart fire alarm system: Generated mode switch table.} 
    \label{fig:modeTable-sfa}
\end{figure}

Figure~\ref{fig:modeTable-sfa} shows that the handler use cases \texttt{ReconnectedToNetwork} and \texttt{ReconnectFDToNetwork} are responsible for returning the system to a normal mode of operation from \texttt{NoInternet} and \texttt{NoAlert} emergency modes. As we discussed earlier, \texttt{ReconnectedToNetwork} is a crucial handler, and this table indicates that successful completion of this handling is required to revert back to normal mode. Thus, this information must be taken into consideration at the design phase. The mode summary table (Table~\ref{modeSummary-sfa}) lists the services available in each mode and aids in understanding if the services in a mode satisfy the minimum requirement of the system. For example, user cannot perform a test on the system or perform maintenance in the \texttt{NoInternet} mode, which does not hamper the ability of the system to detect fire and sound alarm.

%% file: comparison.tex
\section{Comparative Analysis}
\label{sec:evaluation}
We used the smart store system and smart fire alarm system case studies to demonstrate our work. To further assess the effectiveness and usefulness of UCM4IoT as a requirements development approach, we carried out an informal comparative evaluation. Specifically, we selected two approaches for this purpose: 1) the standard use case approach~\cite{fondue2000}, because this is the most commonly used technique for use case-based modelling, and 2) exceptional use cases~\cite{smkd2005}, due to the level of support for exception handling, which exceeds other existing approaches. Use cases for both systems were developed based on the same problem statement using the standard approach, exceptional use cases, and UCM4IoT. 
The goal was to analyze whether UCM4IoT can assist in discovering more system functionalities and exceptions than existing approaches, hence leading to a more correct and complete specification. 
We compared the approaches on the basis of four criteria: exceptions identified, core functionality identified, recovery measures specified, and actors identified. 
Taking these aspects into consideration, we posed the following evaluation questions (EQ) to compare and assess.

 \begin{itemize}
 \item EQ1: Does eliciting requirements with UCM4IoT help in discovering a higher number of (exceptional) situations over exceptional use cases and the standard process, hence leading to a more complete requirements specification?
 \item EQ2a: Does eliciting requirements with UCM4IoT help in revealing and defining core system functionalities that are not identified with exceptional use cases and the standard approach, hence leading to a more complete requirements specification?
 \item EQ2b: Does eliciting requirements with UCM4IoT help in revealing and defining system functionalities, in particular, recovery measures, that are not identified with exceptional use cases and the standard approach, hence leading to a more complete requirements specification? 
\item EQ3: Are we able to discover additional actors that were not identified with standard and exceptional use cases?
  \end{itemize}

Our findings for each of the two case studies are presented here. 

\subsection{Smart Store Case Study}

Since tool support is only available for the UCM4IoT language, we initially documented all three use case models without using the tool. 

 \textbf{EQ1:} Number of exceptions increased by 83.3\% from exceptional use cases to UCM4IoT. We have identified 11 exceptions with UCM4IoT, and 6 exceptions with exceptional use cases. No exception was discovered while developing use cases with the standard process as it does not have the concept of exceptions, nor does the approach enable discovery of such exceptional situations.

\textbf{EQ2a:} UCM4IoT increased the interaction steps in use case models by 15.9\% than exceptional use cases, and 30.8\% than the standard approach. In the standard use case models, there were 39 interactions steps, in exceptional use cases 44, and in UCM4IoT 51. We did not consider interactions steps in handlers for this investigation. UCM4IoT helped to identify several core functionalities of the system that did not come up while eliciting requirements with standard and exceptional use cases process. UCM4IoT ensures that  all the actors participating in achieving a goal must have interaction steps with the system, and the different types of exceptions also led to specification of more interactions that otherwise would be easily missed. 

\textbf{EQ2b:} With UCM4IoT, we defined recovery measures to handle the discovered exceptions. In the process of writing standard use cases, we developed 21 use cases. However, we did not identify functionalities such as handling fire hazard, holding payment in case of payment service issues, requesting camera to take more images if the system is unable to recognize a user, automatically notifying service person in case of a hardware failure, etc.
    There is a significant increase of 166.7\% in defining the recovery measures with UCM4IoT than exceptional use cases. 
    We defined eight handlers with UCM4IoT, whereas there are only three handlers defined with exceptional use cases, and none with standard use cases.
    
 \textbf{EQ3:} New functionalities of the system defined by UCM4IoT introduced new actors. The classification of actors also helped to identify new actors that were not discovered earlier. We identified 30.8\% more actors with UCM4IoT than exceptional use cases, and 70\% new actors than the standard process. Four new exceptional actors, namely \texttt{PhysicalEntity::EmergencyExit}, \texttt{PhysicalEntity::PoliceStation}, \texttt{Device::AttackAlertSwitch}, \texttt{Software::FireDetectionSystem} are discovered with UCM4IoT that were not identified with exceptional use cases. 
 
 In the standard approach, the smart shelf was added as an actor - no sensors or readers part of the smart shelf were identified as actors in this phase. Subsequently, additional actors constituting the smart shelf (\texttt{Device::WeightSensor}, \texttt{Device::PressureSensor}, \texttt{Device::TagReader}) as well as the \texttt{Human::ServicePerson} actor were added to the UCM4IoT use cases.

\subsection{Smart Fire Alarm Case Study}
  
\textbf{EQ1:} Applying the UCM4IoT approach on the Smart Fire Alarm System resulted in a 50\% improvement in the number of exceptions identified compared to the exceptional use cases approach. The total number of exceptions found using exceptional use cases was 12, while the total for UCM4IoT was 18. New exceptions were discovered in 6 different use cases. In the \texttt{UseSmartFireAlarmSystem} use case, two exceptions were found using exceptional use cases, and three exceptions were found using UCM4IoT. Meanwhile, in the \texttt{TestComponents} use case, there were no exceptions found using exceptional use cases, while one exception was discovered using UCM4IOT. In the \texttt{Open Emergency Door} use case, one exception was found using exceptional use cases, and two exceptions were found using UCM4IOT. In the \texttt{NotifySprinklerSystem} and \texttt{AlertFireDepartment} use cases,  there was one exception found using exceptional use cases, while two exceptions were discovered using UCM4IOT. Finally, in the \texttt{Change Display Color} use case, one exception was found using exceptional use cases, and two exceptions were revealed with the help of UCM4IOT.

\textbf{EQ2a:} Eliciting requirements for Smart Fire Alarm System with UCM4IoT helped in revealing and defining core system functionalities that were not identified with exceptional use cases. Applying UCM4IoT resulted in a 18.2\% more interaction steps in the use cases than exceptional use cases and 23.8\% with standard process. There were 21 interaction steps for standard, 22 for exceptional use cases while there were 26 interaction steps using UCM4IoT. The increase in interaction steps was caused by the need to accommodate timeouts, new exceptions, missing actors and network communications. In order to handle exceptions, they need to be first detected, so timeouts were added to the UCM4IoT environment. UCM4IoT’s categorization of exceptions into Software, Hardware, Environment and Network exceptions helped visualize the system into these 4 components. When reviewing the use cases there were missing network communications that were not considered when developing with exceptional use cases. 

\textbf{EQ2b}: There was a 18.2\% increase in recovery measures defined as handlers when using the UCM4IoT approach compared to exceptional use cases. Handlers introduced additional functionalities to the system. Two new handlers called \texttt{RunIBSHardwareTest} and \texttt{RunEmergencyDoorHardwareTest} were created to handle the \texttt{BatteryLevelNotFound} and \texttt{DoorNotWorking} exceptions, which were discovered using UCM4IoT. 
The \texttt{DisplayColorRevert} and \texttt{NoInternet} exceptions were handled using existing handlers.

\textbf{EQ3:} We did not discover any new actors with UCM4IoT. Number of actors for smart fire alarm is same for all three approaches.

\subsection{Summary}

\begin{figure}[tbh!]
\centering
 \includegraphics[width=.6\textwidth]{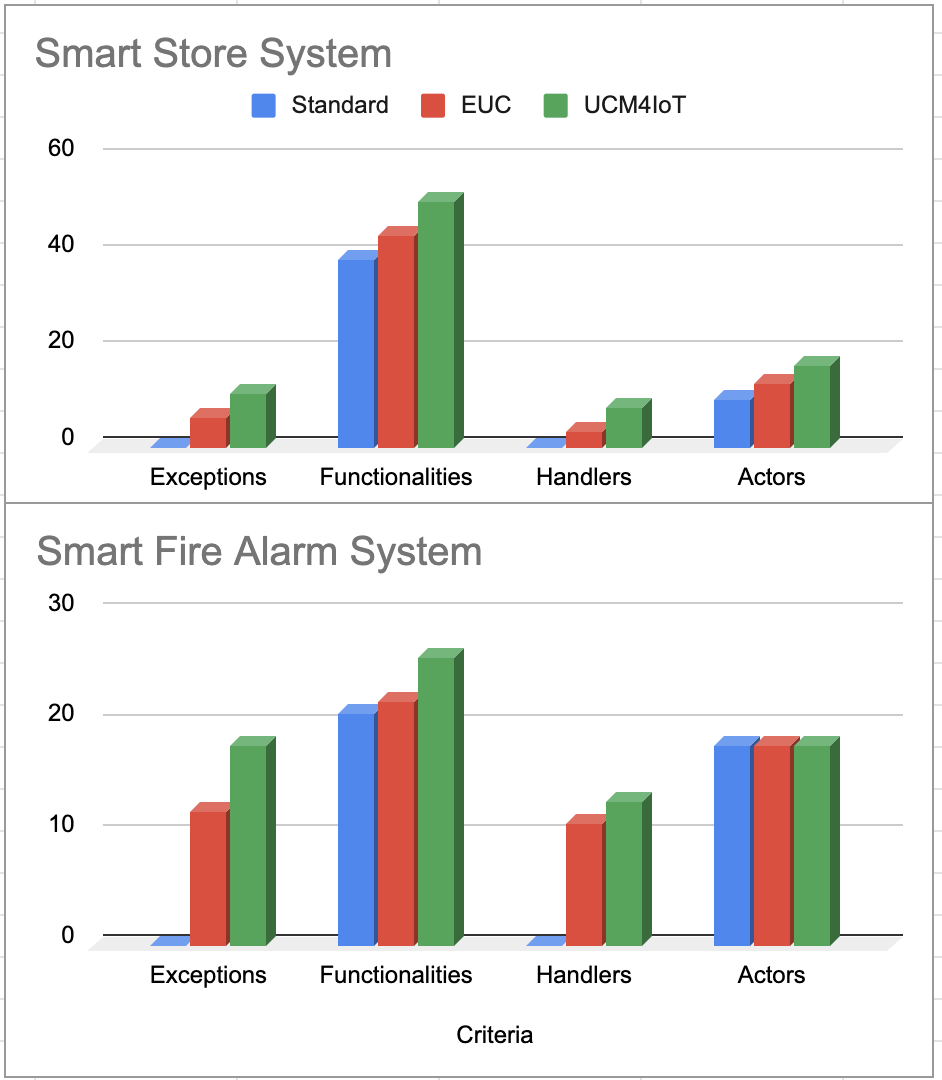}
     \caption{Summary of Findings} 
     \label{fig:comparison}
\end{figure}

Figure~\ref{fig:comparison} summarizes our findings for the smart store and smart fire alarm system case studies.  

Based on both case studies, it is evident that using the domain-specific language, UCM4IoT brought about significant improvements in the requirements specification process of IoT-based systems and led to the development of a more complete and precise requirements model. It allowed an overall increase in the number of exceptional situations discovered and led to identifying missing system functionality (required to handle such situations) as well as core system functionality, thus ensuring that system designers implement mechanisms to handle these exceptions. 

%% file: discussion.tex
\section{Discussion}
\label{sec:discussion}

This work is a first step towards developing a domain-specific requirements development language for IoT. Having IoT-specific language constructs enables modellers to take into consideration the different facets of the domain from the early stages of development. 

UCM4IoT has been applied on several case studies of non-trivial nature selected from different IoT application domains: smart store system, smart fire alarm system, and smart lights system. The latter is not covered in this paper. As illustrated in Section~\ref{sec:evaluation}, there is evidence that UCM4IoT is indeed useful and effective in establishing a more complete and correct requirements model for IoT systems. As the next step in this direction, we intend to apply UCM4IoT on industrial case studies. 

The UCM4IoT models can benefit designers in the later phases of development. 
Identification of hardware exceptions in critical goals could be an indication to have fault tolerance mechanisms (such as hardware redundancy) built into the system at the design phase. Software exceptions can be taken as an indication for incorporating fault prevention mechanisms or additional system functionality by refining the initial specification for critical system operations (for instance, \texttt{SoftwareException::PaymentServiceDown} may lead to a decision further downstream to require customers to provide an alternate payment method). Network exceptions can help in making decisions on the need for offering alternate connectivity options (for instance, \texttt{NetworkException::CustomerUnreachable} may be avoided by making a public Wi-Fi network accessible to all customers inside the smart store).  

Decisions on invoking adaptive mechanisms may be dependent on non-functional requirements (NFRs) or quality constraints associated with a system. These quality attributes may dictate the kind of handler to be used as well as indicate the criticality level of each user or sub-functional goal.  

As future work, we plan on investigating the impact on handlers when taking NFRs into account at the elicitation phase. 

A modelling environment offering syntax-directed editing and validation checks is very useful for documenting correct and complete use cases. The generated exception and handler summary tables provide a global view of the use cases allowing useful insights and inferences to be drawn regarding the requirements. 
Automatically mapping the textual use cases to a use case diagram makes it very easy for modellers to keep the graphical representation consistent with the textual model. Moreover, the UCM4IoT model can be mapped to a (partial) IoT ARM domain model. The use cases can be further used to generate test cases for the system.
We intend to extend the UCM4IoT environment in this direction.

%% file: relatedwork.tex
\section{Related Work}
\label{sec:relatedwork}

In this section, we discuss existing work on use cases as well as requirements modelling for IoT systems.

\subsection{Use Case Modelling}

Sindre and Opdahl \cite{sindre2005eliciting} present a method for eliciting security requirements by extending use cases. Misuse cases only address security risks and and was designed to mitigate potential threats exposed by misusers. 
Savić et al.~\cite{savic2012} propose a DSL, SilabReq, that allows use cases to be defined at different abstraction levels depending on the type of stakeholder. 

Cockburn~\cite{cockburn2001} informally addresses the notion of failure scenarios in standard use cases with alternative situations. 
Amyot~\cite{amyot2001use} proposes use case maps as part of the URN standard for specifying functional requirements and causal relationships between use cases. This approach implicitly addresses the detection of undesirable scenarios. 

Mustafiz et al.~\cite{mustafiz2009drep} propose a requirements engineering process, DREP, to guide developers in considering reliability and safety concerns of reactive systems starting at the use case level. Two types of exceptions, context-affecting and service-related, are used to define exceptions and associated dependability handlers. 
DREP does not come with any tool support for use case modelling.

UCM4IoT is based on the notion of exception handling in use cases proposed in \cite{smkd2005} and \cite{mustafiz2009drep}. UCM4IoT takes this further by proposing IoT-specific concepts in use case modelling along with a dedicated environment for editing, validating and analyzing use cases. 

Maleki et al.~\cite{maleki2019framework} propose a framework for exception discovery and handling in the elicitation phase to be used for developing test cases. Requirements are defined with graphs in three levels of granularity (goals, scenarios, and sub-functions) along with a fault tree to present probable error or failure.  While UCM4IoT can be extended for model-based requirements testing, at this time, it does not cover testing.

While there exists several modelling tools with support for use case diagrams, very few environments are available for textual use case modelling. Visual Use Case (\url{http://www.visualusecase.com/}) supports standard use cases editing and mapping to an activity diagram. UCEd~\cite{some2004-uced} also provide support for textual use case modelling. However, these tools do not cover exceptions or any IoT-related aspects and are not based on MDE techniques. 
The RUCM~\cite{yue2013-rucm} use case modelling approach supports global and bounded alternatives that could be used to define exceptions, but no explicit language-level support is provided for IoT systems. 

To the best of our knowledge, UCM4IoT is the only existing one to provide support for model-driven requirements development with textual use case modelling for IoT systems. 

Our environment supports discovery and documentation of UCM4IoT use cases as well as generation of exception summary information that can be used for further analysis of the requirements. While UCM4IOT currently does not provide integrated support for use case diagrams, we use an extended form of use case diagrams for graphically representing and summarizing our use cases.

\subsection{Requirements Modelling for IoT}

Reggio~\cite{reggio2018} introduces IoTReq, a goal-oriented method for elicitation and specification of IoT system requirements, which uses UML profiles for domain modelling.  
Different types of behaviour are specified including unexpected behaviour or constraints (expressed with a sequence or activity diagram) imposed by the system that need to be made impossible by the IoT system. In comparison, exceptions (as used in our work) are situations which hinder fulfillment of goals. 
Meacham et al.~\cite{meacham2016} propose a requirements modelling approach for IoT systems combining Volere template (for organizing requirements in English), UML use cases and SysML diagrams. Their approach considers exceptional cases in use case diagrams by finding abnormal conditions, but, it does not specify types of exceptions and actors like UCM4IoT. There is no support for elicitation with textual use cases. Although this work targets the IoT domain, their method is generic and does not specify any IoT-specific requirements. 
Sosa-Reyna et al. \cite{sosa2018methodology} propose a methodology to develop IoT systems with a service-oriented approach. They used formal transformation rules to go from one phase to another in their four phase model-based system development methodology. In the first phase of requirements engineering, UML use case and activity diagrams are used to elicit both functional and non-functional requirements. They also presented a metamodel to generate smart vehicle applications following their proposed methodology. 

Silva et al. \cite{silva2019requirements} proposed to modify the ISO and
IEEE standards IEC/IEEE 12207:2017 of requirements engineering for IoT systems. They defined the requirements engineering of IoT in three sub-processes: i) process scope definition, ii) IoT system definition, and iii) IoT system requirements definition. Each sub-process then follows the ISO and IEEE standard process. This work presents a process for requirements engineering, but unlike our work, it does not propose any domain-specific language or environment to facilitate the elicitation and specification of IoT system requirements.

A comparison of the related work is presented in Table~\ref{table:relatedwork}. Currently, limited methods, techniques, or tools are available for requirements engineering of IoT systems. Most of the existing work focus on design and development phases.  However, to minimize errors and changes in the downstream phases, a well-defined set of requirements is essential. UCM4IoT will help fill this gap, ensuring IoT requirements including exceptional and adaptive behaviour are discovered and documented from the early stages of software development.

\begin{table}
\fontsize{7}{8}\selectfont
\centering
\caption{Comparison of existing work (supports (\checkmark), does not support (x), unknown/unclear (-)}
\label{table:relatedwork}
\begin{tabular}{|p{1.6cm}|p{2cm}|p{.8cm}|p{3.2cm}|p{.9cm}|p{.9cm}|p{.9cm}|}

\hline

Approach & Tool Support & IoT Support  & Analysis Support & Elicita-tion &  Specifi-cation &  Valida-tion \\
\hline
     Sindre and Opdahl \cite{sindre2005eliciting}      & \vspace{1mm}  \large   x       &   \vspace{1mm} \large   x            &      Threat analysis from the misuse cases  & \vspace{1mm}  \large \checkmark & \vspace{1mm} \large \checkmark  & \vspace{1mm} \large x     \\\hline
     
     Savić et al. \cite{savic2012}         & Model to Model transformation, SilabReq models to UML models   &   \vspace{1mm}  \large  x      &  Requirements analysis from different abstraction levels &\vspace{1mm}  \large \checkmark  & \vspace{1mm} \large \checkmark &  \vspace{1mm} \large  x  \\\hline
     
      Cockburn \cite{cockburn2001}      &     \vspace{1mm} \large  x           &    \vspace{1mm} \large  x               & \vspace{1mm} \large  x               & \vspace{1mm} \large  \checkmark   &  \vspace{1mm} \large  x    &  \vspace{1mm} \large  x   \\\hline
        Amyot \cite{amyot2001use}       &  UCM Navigator for drawing use case maps  &  \vspace{1mm} \large  x            &  Use case maps to minimize gap between requirements and design       & \vspace{1mm} \large  \checkmark  &  \vspace{1mm} \large  \checkmark   & \vspace{1mm} \large x     \\\hline
          Mustafiz et al. \cite{mustafiz2009drep}        &       \vspace{1mm} \large  x     &    \vspace{1mm} \large x          &        \vspace{1mm} \large  x   & \vspace{1mm} \large  \checkmark   &    \vspace{1mm} \large  x      &      \vspace{1mm} \large  x     \\\hline
          Maleki \cite{maleki2019framework}       &    \vspace{1mm} \large  x         &       \vspace{1mm} \large  x       &              Supports to derive software critical path    &\vspace{1mm} \large  \checkmark &\vspace{1mm} \large  \checkmark   &  \vspace{1mm} \large  x  \\\hline \hline
          Reggio \cite{reggio2018} &   \vspace{1mm} \large  x             &   \vspace{1mm} \large  \checkmark            &        Supports finding unexpected behaviour and constraints to goals & \vspace{1mm} \large  \checkmark  &    \vspace{1mm} \large  \checkmark      &   \vspace{1mm} \large  x  \\\hline
         Meacham \cite{meacham2016}  &    \vspace{1mm} \large  x            &       \vspace{1mm} \large  \checkmark       &      Abnormal condition analysis from use case diagram  & \vspace{1mm} \large  \checkmark & \vspace{1mm} \large  \checkmark &     \vspace{1mm} \large  x      \\\hline
      Sosa-Reyna et al. \cite{sosa2018methodology}     &         \vspace{1mm} \large  x      &    \vspace{1mm} \large  \checkmark                   & \vspace{1mm} \large x & \vspace{1mm} \large  \checkmark & \vspace{1mm} \large  \checkmark & \vspace{1mm} \large  x\\\hline
Silva et al.\cite{silva2019requirements}  &    \vspace{1mm} \large  x &    \vspace{1mm} \large  \checkmark   & \vspace{1mm}\large - &  \vspace{1mm} \large  \checkmark & \vspace{1mm} \large  \checkmark & \vspace{1mm} \large  \checkmark\\ \hline 

\end{tabular}
\end{table}

%% file: conclusion.tex
\section{Conclusion}
\label{sec:conclusion}

We have proposed a requirements development language, UCM4IoT, catered for IoT systems. Our textual use case language provides IoT-specific language constructs for discovering the different types of actors and interactions participating in an IoT system. Our approach enables exceptional scenarios associated with the different facets of IoT (hardware, software, network, and environment) to be identified during requirements elicitation. This is followed by discovering adaptive system behaviour as handling mechanisms. UCM4IoT allows specification of these potential exceptions in use cases with supporting handler use cases to document exceptional system interactions. UCM4IoT also supports specification of different modes of operation, including the available services in each mode. 
 A textual modelling environment for UCM4IoT has been developed to assist modellers in writing unambiguous and complete use cases. The tool enables syntax highlighting, type-checking, cross-referencing, and global validation of use cases. Support for generation of exception and mode summary information is also provided to enable exploration and static analysis of the use cases. We also extended the UML use case diagram with IoT-specific elements that align with our textual language.

Our approach is demonstrated and evaluated with two IoT applications, a smart store system and a smart fire alarm system, encompassing hardware, software, human users, and physical entities all working in coordination to provide the system functionalities and fulfill the user goals. 

As future work, we plan on integrating NFRs in UCM4IoT and providing support for dynamic analysis of requirements.